\documentclass[12pt]{article}
\usepackage[dvips]{graphicx}
\usepackage{epsfig,cite}
\usepackage {amsmath}
\usepackage{color}
\usepackage{rotating}
\usepackage{amssymb}
\usepackage{slashed}
\usepackage[bookmarks]{hyperref}
\include{epsf}
\newcount\timecount
\newcount\hours \newcount\minutes  \newcount\temp \newcount\pmhours
\hours = \time
\divide\hours by 60
\temp = \hours
\multiply\temp by 60
\minutes = \time
\advance\minutes by -\temp
\def\hour{\the\hours}
\def\minute{\ifnum\minutes<10 0\the\minutes
            \else\the\minutes\fi}
\def\clock{
\ifnum\hours=0 12:\minute\ AM
\else\ifnum\hours<12 \hour:\minute\ AM
      \else\ifnum\hours=12 12:\minute\ PM
            \else\ifnum\hours>12
                 \pmhours=\hours
                 \advance\pmhours by -12
                 \the\pmhours:\minute\ PM
                 \fi
            \fi
      \fi
\fi
}

\def\monthname{\relax\ifcase\month 0/\or January\or February\or
   March\or April\or May\or June\or July\or August\or September\or
   October\or November\or December\else\number\month/\fi}

\def\bold#1{\setbox0=\hbox{$#1$}%
     \kern-.025em\copy0\kern-\wd0
     \kern.05em\copy0\kern-\wd0
     \kern-.025em\raise.0433em\box0 }

\def\beq{\begin{equation}}
\def\eeq{\end{equation}}

\def\ga{\mathrel{\raise.3ex\hbox{$>$\kern-.75em\lower1ex\hbox{$\sim$}}}}
\def\la{\mathrel{\raise.3ex\hbox{$<$\kern-.75em\lower1ex\hbox{$\sim$}}}}
\def\gev{{\rm \, Ge\kern-0.125em V}}
\def\tev{{\rm \, Te\kern-0.125em V}}
\def\gyr{{\rm \, G\kern-0.125em yr}}

\def\gappeq{\mathrel{\rlap {\raise.5ex\hbox{$>$}}
{\lower.5ex\hbox{$\sim$}}}}
\def\lappeq{\mathrel{\rlap{\raise.5ex\hbox{$<$}}
{\lower.5ex\hbox{$\sim$}}}}
\def\Toprel#1\over#2{\mathrel{\mathop{#2}\limits^{#1}}}

\def\m12{m_{1\!/2}}

\def\bea{\begin{eqnarray}}
\def\eea{\end{eqnarray}}

\def\beqar{\begin{eqnarray}}
\def\eeqar{\end{eqnarray}}

\usepackage{lscape}
\usepackage[dvips]{graphicx}

\def\beq{\begin{equation}}
\def\eeq{\end{equation}}

\makeindex

\begin{document}

\begin{titlepage}
\pagestyle{empty}
\rightline{KCL-PH-TH/2013-05, LCTS/2013-02, CERN-PH-TH/2013-015}
\vskip +0.4in
\begin{center}
{\large {\bf LHC Missing-Transverse-Energy Constraints\\
\vskip 0.1in
 on Models with Universal Extra Dimensions}}

\end{center}
\begin{center}
\vskip +0.3in
{\bf Giacomo~Cacciapaglia}$^{1,2}$, {\bf Aldo Deandrea}$^{1}$, {\bf John~Ellis}$^{2,3}$, \\
\vskip 0.1in
{\bf Jad Marrouche}$^{4}$ and {\bf Luca Panizzi}$^{5}$ \\
\vskip 0.2in
{\small {\it
$^1${Universit\'e de Lyon, F-69622 Lyon, France; Universit\'e Lyon 1, Villeurbanne;
CNRS/IN2P3, UMR5822, Institut de Physique Nucl\'aire de Lyon,
F-69622 Villeurbanne Cedex, France}\\ 
$^2${Theoretical Particle Physics and Cosmology Group, Department of Physics, King's College London, London~WC2R 2LS, UK}\\
$^3${TH Division, Physics Department, CERN, CH-1211 Geneva 23, Switzerland}\\
$^4${High Energy Physics Group, Blackett Laboratory, Imperial College, Prince Consort Road, London SW7 2AZ, UK}} \\
$^5${\it School of Physics and Astronomy, University of Southampton,\\ Highfield, Southampton SO17 1BJ, UK}
}
\vskip 0.2in
{\bf Abstract}
\end{center}
{\small
We consider the performance of the ATLAS and CMS searches for events with missing transverse
energy, which were originally motivated by supersymmetry, in constraining extensions of the
Standard Model based on extra dimensions,  in which the mass differences between
recurrences at the same level are generically smaller than the mass hierarchies in 
typical supersymmetric models. We consider first a toy model with
pair-production of a single vector-like quark $U_1$ 
decaying into a spin-zero stable particle $A_1$ and jet, exploring the sensitivity of the CMS $\alpha_T$ and ATLAS $m_{eff}$ analysis to
$M_{U_1}$ and the $U_1 - A_1$ mass difference. For this purpose we use versions of the {\tt Delphes}
generic detector simulation with CMS and ATLAS cards, which have been shown to reproduce
the published results of CMS and ATLAS searches for supersymmetry. We then explore the
sensitivity of these searches to a specific model with two universal extra dimensions, whose signal is dominated by the pair production of quark recurrences, including searches with leptons.
We find that the LHC
searches have greater sensitivity to this more realistic model, due partly to the contributions of
signatures with leptons, 
and partly to events with large missing transverse energy generated by
the decays of higher-level Kaluza-Klein recurrences. We find that the CMS $\alpha_T$ analysis
with $\sim 5$/fb of data at 7~TeV excludes a recurrence scale of 600~GeV at a confidence
level above 99\%, increasing to 99.9\% when combined with the CMS single-lepton search,
whereas a recurrence scale of 700~GeV is disfavoured at the 72\% confidence level.
}


\vfill
\leftline{February 2013}
\end{titlepage}

\tableofcontents

\section{Introduction}

One of the most powerful tools in the LHC quest for physics beyond the
Standard Model is the search for events with missing transverse energy
(MET), such as might be carried away by invisible dark matter
particles. The prototypes for models predicting MET events have been
provided by supersymmetry with R-parity, 
and the published CMS and ATLAS
searches for MET events have been designed largely with supersymmetric
models in mind~\cite{LHCsearches}. They are particularly sensitive to
models with a large hierarchy between the masses of
strongly-interacting sparticles, which are the most copiously produced
at the LHC, and the dark matter particle, which has often been assumed
to be the lightest neutralino. However, the sensitivities of some of
these searches to supersymmetric models with smaller mass hierarchies
have also been explored~\cite{LeCompte:2011fh,Dreiner:2012gx,Bhattacherjee:2012mz}.

There are, however, many other classes of extensions of the Standard
Model that also predict MET events, some with different generic
features from the supersymmetric models with a large hierarchy of
masses for which the initial LHC MET searches were optimised.
Examples are models with universal extra dimensions (UEDs)
compactified in a space with characteristic scale $R \sim
1$/TeV~\cite{Antoniadis:1990ew,Appelquist:2000nn}. Since this
compactification scale is much smaller than the typical unification
scale in supersymmetric models, the masses of the Kaluza-Klein
recurrences of known particles are renormalised by relatively small
amounts, and the mass differences between, e.g., the recurrences of
strongly- and weakly-interacting particles at the same level are much
less than would be the mass differences between strongly- and
weakly-interacting sparticles in models with high unification
scales. If such an extra-dimensional model has an exactly-conserved
Kaluza-Klein K-parity~\cite{Appelquist:2000nn,LKP}, the lightest
recurrence is stable, and hence present as a dark-matter relic from
the Big Bang and liable to be produced in MET events at the LHC.  A
related class of extensions includes Little Higgs
models~\cite{ArkaniHamed:2002qx,ArkaniHamed:2002qy}, which can be
thought of as minimal four-dimensional versions of deconstructed extra
dimensions~\cite{ArkaniHamed:2001ca}.  The inclusion of a
T-parity~\cite{Low:2004xc} can stabilise the lightest neutral partner
of the gauge bosons and thus also offer a dark matter candidate.

The sensitivity of the initial ATLAS $m_{eff}$ search to
supersymmetric models with different mass hierarchies was studied
in~\cite{LeCompte:2011fh}, and the sensitivity to the soft QCD
jet-matching procedure and the sensitivities of other searches by both
ATLAS and CMS ($\alpha_T$, $m_{T2}$, razor variables and monojet) have been studied
in~\cite{Dreiner:2012gx}.  In both cases, simplified models containing
only near-degenerate gluinos, squarks and an LSP have been used.
However, there are important differences between these and
extra-dimensional scenarios. In particular, the spins of the
Kaluza-Klein recurrences are different from those of the hypothetical
supersymmetric partners of Standard Model particles, affecting both
production cross sections and decay patterns.  For example, the decay
chains necessarily include a sizeable fraction of leptons from the
decays of the recurrences of the SU(2) gauge bosons, so searches
containing leptons and MET must also be considered \cite{Bhattacherjee:2010vm,Belyaev:2012ai}
(this is also true for supersymmetric spectra containing light sleptons).
More importantly,
the production and decays of higher-level recurrences, which are even
under the Dark Matter parity, may also make important contributions to
prospective MET signatures.

In this paper we study various aspects of these potential features in
searches for Kaluza-Klein excitations via MET searches at the LHC. Our
first step is to define a minimal simplified toy model comprising a
single excitation of the $u$ quark, $U_1$, that decays exclusively
$U_1 \to u + A_1$, where $A_1$ is a neutral, weakly-interacting scalar
particle.  This toy model should encompass important features of the
signal in realistic UED and Little Higgs models, which is dominated by
the pair production of the lightest K- or T-odd recurrences of the
light quarks.  Here we consider a scalar $A_1$, which is typical of
6-dimensional models, whereas a vector $A_1$ would be expected in
5-dimensional and Little Higgs models.  The toy model is also closer
to the simplified models used to study compressed supersymmetric
spectra, thus allowing us to make a direct comparison of the
acceptances of the searches on the two kinds of signals~\footnote{A
similar toy model has been used in~\cite{Perelstein:2011ds} to study
the reach in Little Higgs models with T-parity: in our study, however,
we focus on a region of parameter space with smaller mass splitting.}.
We study the prospective sensitivities of the searches looking for MET
and jets in the $(M_{U_1}, M_{A_1})$ plane of this toy model, using
variants of the {\tt Delphes} generic detector simulation with CMS and
ATLAS detector cards that have been validated for MET searches in the
contexts of supersymmetric models.  In particular, we use simulated
signal events to analyse the efficiencies of the CMS $\alpha_T$,
monojet, single-lepton (Lp), opposite-sign (OS) and same-sign (SS)
dilepton analyses, and of the ATLAS $m_{eff}$ search using the 2011
data at 7 TeV, which have now been finalised.  For
definiteness, and because it has greater sensitivity than the ATLAS
$m_{eff}$ search and monojet searches, we focus on the CMS $\alpha_T$
analysis, with a view to understanding how to optimise the strategies
for future searches for Kaluza-Klein excitations.

The main goal of this preliminary analysis is to validate our
implementation of the experimental cuts, as a step to extending the
analysis to a complete and realistic UED-type model.  Several
possibilities for such models have been proposed in the literature: in
the present work we focus on the simplest possible model with a pair
of universal extra dimensions, which is the minimal model with an
exact K-parity proposed in~\cite{Cacciapaglia:2009pa}.  The main
advantage of this model is that it has an exact symmetry, which is a
property of the geometry of the compact space and does not require any
additional assumption on the interactions localised at the singular
points of the space.  We chose this model also for the following two
reasons: the loop-induced mass splitting in each tier of recurrences
is the smallest among the proposed models, thus offering an ideal
arena to probe the sensitivity to compressed spectra. Also, the
preferred mass range indicated by the Dark Matter abundance is $700 <
m_{KK} < 1000$ GeV~\cite{Arbey:2012ke}, which implies large production
rates at the LHC and the possibility of probing the preferred masses
with the present data.  In contrast, UED models in five dimensions
prefer larger values of $m_{KK} \sim 1.5$ TeV~\cite{LKP}.

We have implemented the model in {\tt
MadGraph5}~\cite{Alwall:2011uj}, and performed an
analysis of its parameter space, including the full pattern of decays
of the lowest even and odd tiers of excitations. The production cross
section is dominated by the pair production of the lightest K-odd
recursions of the light quarks, which roughly correspond to the toy model,
while other contributions, specific to extra-dimensional models, come
from the K-odd scalar gluon and the higher K-even recurrences.  We use
the complete simulation to check first how closely the events from the
K-odd quark pair production are described by the toy model.  We then
compare the full set of channels, and study in detail the effect of
all the searches, including MET searches involving leptons, in order
to identify the production channels and searches that provide the main
contributions to the combined bound.  This exercise makes it possible
to compare the bound with those obtained from other searches without
MET, primarily the search for resonant dileptons coming from the decay of even recurrences of the neutral gauge bosons~\cite{Cacciapaglia:2012dy}. It also provides some
indications how to modify the supersymmetry searches to
increase the sensitivity to UED-type models.
In this paper we only include the final searches based on the 2011 dataset at 7 TeV, and we postpone the analysis of the preliminary 2012 results and 
a study of the performance of alternative searches in a follow-up publication.
We expect the inclusion of the 8 TeV data to certainly improve on the bounds, however without changing significantly the conclusions of this paper.

The layout of the paper is as follows. In Section~2 we introduce the toy
model with a single $u$ excitation $U_1$ and a stable neutral scalar $A_1$, 
describe our implementations of the
CMS $\alpha_T$ and ATLAS $m_{eff}$ 
MET analyses using {\tt Delphes},
and go on to discuss the sensitivities of these searches in the $(M_{U_1}, M_{A_1})$
plane that characterises the toy model. Then, in Section~3 we review the simplest
model with two UEDs, 
and discuss the
sensitivity of the CMS MET searches, including single-
and dilepton searches, to this more complete scenario. We find that the LHC
searches are more sensitive to this more realistic model, thanks partly to 
the decays of higher-level Kaluza-Klein recurrences that generate events with large missing transverse energy, and partly to 
the contribution of the single-lepton
signature. Based on these results, in Section~4 we make suggestions how the present CMS and ATLAS
MET search strategies could be modified to be more sensitive to such a UED scenario,
and in Section~5 we present a summary and conclusions.

\section{Getting Started}

\subsection{The Toy Model}

We describe now the toy model we use to exemplify the features we seek
to explore in this paper.  To keep things as minimal as possible, we
consider only the recurrences of the light quarks: in generic UED
models, they are nearly degenerate because their universal mass is
fixed by the size of the extra compact dimensions, with only small
splittings generated by loop corrections.  In Little Higgs models, the
masses are arbitrary, but degenerate spectra are preferred in order to
minimise corrections to flavour observables.

The most minimal realisation of the toy model consists of only two new
particles: a heavy up-type quark, $U_1$, and a neutral stable scalar,
$A_1$.  In the following, we assume that $M_{U_1}>M_{A_1}$, and that
the only allowed decay channel for the new quark is $U_1\to u A_1$
with coupling $c_{U_1A_1u}=2e/3c_w^2$.
The couplings has been chosen to match the coupling of the hypercharge gauge boson, as one usually obtains in realistic models.
However, the precise value of the coupling is not relevant for the study, as long as the decay is prompt 
and the width small enough compared to the experimental accuracy on the MET and jet energy measurements.
The neutral scalar cannot decay into any lighter
state, and is therefore a Dark Matter candidate, invisible to the detector.  The only free
parameters of the toy model are the heavy quark mass $M_{U_1}$ and the
mass splitting $\Delta M = M_{U_1} - M_{A_1}$.  In the following we
will be interested in near-degenerate spectra, and we consider mass
splitting below $100$ GeV~\footnote{This toy model can be considered
as a UED equivalent of the simplified supersymmetric model that only
contains degenerate squarks and an LSP, with gluinos and non-coloured
sparticles assumed to be decoupled.}.

\subsection{Simulation Details} \label{sec:simulation}

The processes for pair production of new particles in the toy model
can be grouped into three classes:
\begin{eqnarray}
p p \to \left\{\begin{array}{c} Q_1 Q_1 \\ Q_1 A_1 \\ A_1 A_1
\end{array}\right\} + n~{\rm jets} \quad~{\rm where}~\quad
Q_1=U_1,\bar U_1\,,
\end{eqnarray}
with subsequent decay of the heavy quark $U_1$ into the invisible scalar $A_1$ and
a quark $u$.
The $Q Q$ class is by far the dominant one, because it includes the
QCD production of $U_1 \bar{U}_1$, which depends only on the mass of
the heavy quark.  Another relevant production channel in this class is
$U_1 U_1$, mediated by t-channel $A_1$ exchange: this process may give
a sizeable contribution at large $M_{U_1}$ masses, because the
suppression due to the electroweak couplings would be compensated by
an enhancement of the up-quark PDFs.  However, this process also
depends on the model-dependent $U_1 A_1 u$ coupling: for this reason,
in the following we consider only the QCD pair production and neglect
all the electroweak processes.  The events we are interested in are
triggered by the presence of high-$p_T$ jets, which can be originated
either by the decays of the heavy quark $U_1$, when the mass splitting
is large or the heavy quark is produced with a significant boost, or
by additional QCD jets due to initial- or final-state radiation (ISR
or FSR).  For the latter reason, we need to incorporate events with
additional hard QCD jets and carefully take into account the matching
of the hard jet generation with the soft QCD jet incorporation.

\begin{table}
\centering\begin{tabular}{|c||c|c|c|c|c|c|c|c|}
\hline
Mass (GeV) & 250 & 300 & 350 & 400 & 450 & 500 & 550 & 600 \\
\hline
Cross-section (pb) & 22.1  & 7.83  & 3.14  & 1.38  & 0.65 & 0.324 & 0.168  & 0.091 \\
\hline
\end{tabular}
\caption{\it Cross sections for $U_1\bar U_1$ production at 7~TeV calculated in the toy model at NLO-NNLL~\cite{Cacciari:2011hy,Cacciari_ttbar}.}
\label{sigmastoymodel}
\end{table}

The toy model signal has been simulated with {\tt
MadGraph5}~\cite{Alwall:2011uj}, interfaced with {\tt
PYTHIA}~\cite{Sjostrand:2006za} for parton showering and
hadronisation, and {\tt Delphes}\cite{Ovyn:2009tx} for the detector
simulation of both ATLAS and CMS.  The simulation has been conducted
scanning over $M_{U_1}$ and $\Delta M$ in the following ranges:
\begin{eqnarray}
 M_{U_1} &=& \{250,600\}~\mbox{GeV} \quad \textrm{in steps
 of}~50~\mbox{GeV}\,; \nonumber \\ \Delta M &=&
 5,20,35,50,75,100~\mbox{GeV}\,. \nonumber
\end{eqnarray}
The cross sections calculated at NLO-NNLL for $U_1\bar U_1$ production
at 7~TeV for the different values of $M_{U_1}$ are shown in
Table~\ref{sigmastoymodel}.  We generated 300,000 events for each of
these choices of $M_{U_1}$ and $\Delta M$ to ensure our equivalent
integrated luminosity was larger than that collected at $7$~TeV by the
ATLAS and CMS experiments.  The region with smaller mass splitting has
been explored in more detail because, in generic UED models, particles
in the same tier are almost degenerate and the small mass splittings
arise only at loop level.

Initial and final-state radiation (ISR and FSR) have been included
both at the level of matrix-element generation and the parton shower,
and the matching has been implemented using the ``shower-$k_T$''
scheme~\cite{Alwall:2008qv} encoded in {\tt Madgraph5}: with this
choice, the number of rejected events from {\tt Madgraph} is low,
enhancing the number of available events.  The matching parameters
$xqcut$ (equal to $qcut$ in the ``shower-$k_T$'' scheme) has been
tuned for every $M_{U_1}-\Delta M$ configuration in order to optimise
the matching procedure for each process.  Kinematical cuts on jets and
missing energy have been set to minimal values at the generator level;
tighter cuts have been applied after the detector simulation, at the
level of the analysis, following the cuts used by the ATLAS and CMS
collaborations. More details are provided in the next sections.

\subsection{Implementations of {\tt Delphes} and Kinematic Distributions in the Toy Model}
\label{sec:analysis}

We use for our analysis versions of the {\tt Delphes} generic detector
simulation package~\cite{Ovyn:2009tx} with `cards' that emulate the
performances of the ATLAS and CMS detectors. Our implementations of
{\tt Delphes} had previously been validated in analyses of
supersymmetry searches using MET signatures~\cite{MC8}.  The first
step in that validation was to use {\tt
SoftSUSY}~\cite{Allanach:2001kg} to generate the spectra for various
CMSSM points, then {\tt PYTHIA}~\cite{Sjostrand:2006za} to generate event
samples for each of these points, followed by {\tt
Delphes} with the ATLAS or CMS detector `card'.  We
then implemented the experimental cuts obtained
from~\cite{LHCsearches}, and calculated corresponding confidence
levels for these points.  As shown in~\cite{MC8} for the case of the
ATLAS MET analysis with 5~fb$^{-1}$ of data at 7~TeV~\cite{ATLAS5fb},
this procedure reproduced quite accurately the 95\% CL bound quoted by
ATLAS for the CMSSM, and we have performed a similar validation for
the CMS MET search. The success of these {\tt Delphes} validations in
another extension of the Standard Model with a complex spectrum of
massive states reinforces the expectation that {\tt Delphes}
simulations are suitable for models whose kinematic distributions lie
well within the analysis search regions, such as for the
extra-dimensional models considered here. We do not make any attempt
to simulate SM background predictions, for which this is not the case.

\begin{figure}
\centering\epsfig{file=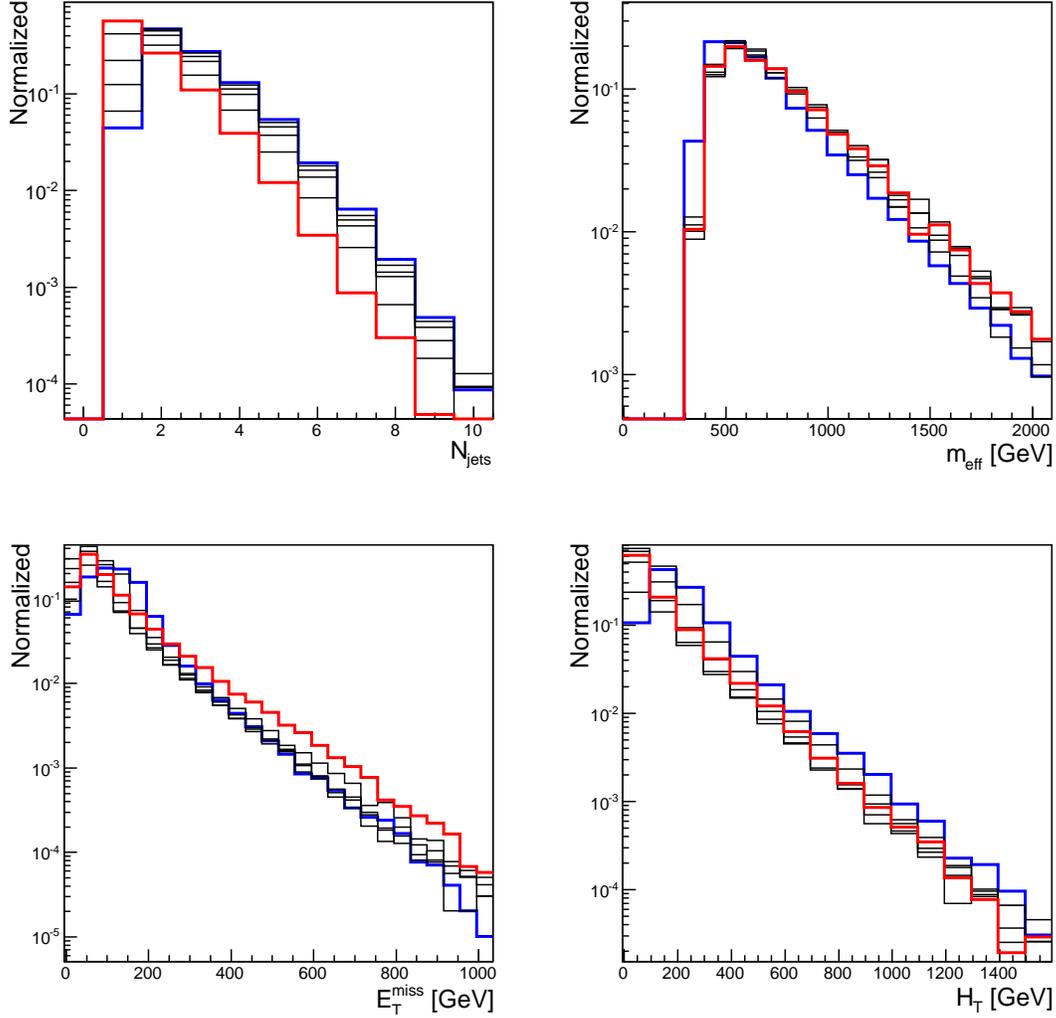,width=.9\textwidth}
\caption{\it Various kinematical distributions ($N_{jets}$, $m_{\rm eff}$, $E_T^{\rm miss}$ and $H_T$) for the ATLAS 5/fb analysis for
$M_{U_1} = 600$~GeV and varying $M_{U_1} - M_{A_1}$ from 100~GeV (blue) to 5~GeV (red). 
All the distributions are normalised to unity.}
\label{fig:ATLASdistributions400} 
\end{figure}

The ATLAS search~\cite{ATLAS5fb} consists of 11 overlapping signal
regions, which differ in the numbers of jets required and in the cuts
on $m_{eff}$ that were applied.  In all cases, events with an electron
(muon) with $p_T > 20$ GeV ($10$ GeV) are vetoed.  We have implemented
all these 11 regions and retained for each choice of the toy model
parameters the region that sets the most stringent bound.  The
11 regions have very different sensitivities to the toy model, and the
two regions with the highest sensitivities (largest signal yields) are
in general A$^\prime$ and C(900). Both of these require missing transverse energy
$E_T^{miss} > 160$~GeV,
two jets with $p_T(j_1, j_2) > 130, 60$~GeV respectively and $E_T^{miss}/m_{eff} >
0.25$.  In addition, A$^\prime$ requires $m_{eff} > 1200$~GeV whereas C(900)
requires two more jets with $p_T(j_3, j_4) > 60$~GeV and $m_{eff} >
900$~GeV.
Fig.~\ref{fig:ATLASdistributions400} displays various kinematical
distributions (normalised to unity) for the ATLAS 5/fb analysis for
fixed quark mass $M_{U_1} = 600$~GeV and varying the splitting
$M_{U_1} - M_{A_1}$ from 100~GeV (blue) to 5~GeV (red). We see in the
panels significant variations in the $N_{jets}$, and hence $H_T$ and
$E_T^{Miss}$, distributions as $M_{U_1} - M_{A_1}$ varies.  For larger
splitting, the average number of jets reconstructed above the
respective $p_T$ threshold increases due to the fact that the jets
from the decay have larger $p_T$ whereas, for very small splitting,
the high-$p_T$ jets originate mainly from QCD radiation.
Also shown in the top right panel is the $m_{eff}$ distribution after
the ATLAS event selection for region A$^\prime$, again normalised to unity.
We observe that the $m_{eff}$ distribution is rather insensitive to
the mass splitting, and similar conclusions apply to
other values of the parameters of the toy model.

\begin{figure}
\centering\epsfig{file=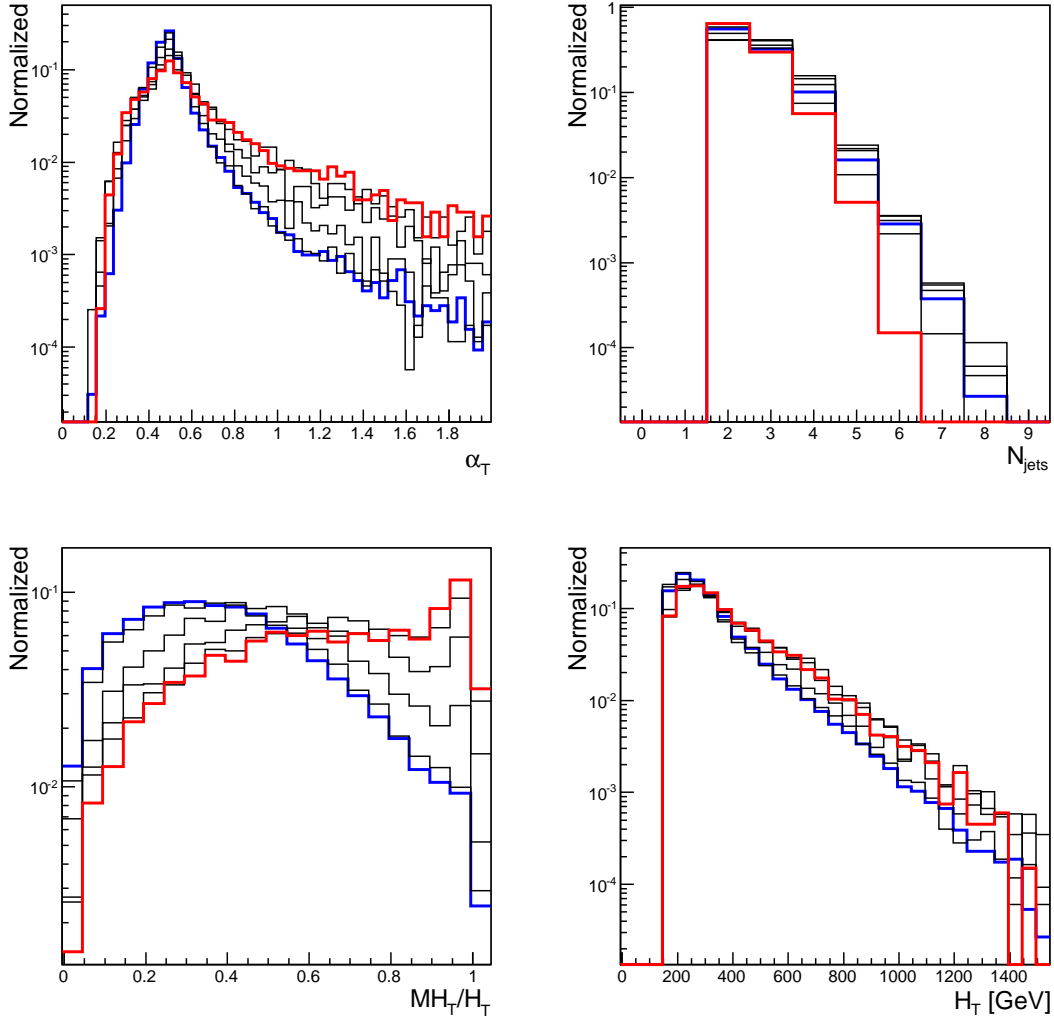,width=.9\textwidth}
\caption{\it Various kinematical distributions ($\alpha_T$, $N_{\rm jets}$, $MH_T/H_T$ and $H_T$) for the CMS $\alpha_T$ analysis for
$M_{U_1} = 600$~GeV and varying $M_{U_1} - M_{A_1}$ from 100~GeV (blue) to 5~GeV (red). 
All the distributions are normalised to unity.}
\label{fig:CMSdistributions400} 
\end{figure}

The CMS collaboration has implemented various searches based on
different kinematical variables: $m_{T2}$~\cite{CMSmT2}, razor
variables~\cite{CMSrazor} and $\alpha_T$~\cite{Chatrchyan:2011zy},
which have similar sensitivities for the CMSSM.  Here we focus on the
7~TeV $\alpha_T$ based search using 5 fb$^{-1}$, because its
documentation provides the details needed for
our analysis.  The kinematic variable $\alpha_T$, defined as 
\beq
\alpha_T = \frac{p_T (j_2)}{M_{jj}} = \frac{p_T (j_2)}{\sqrt{H_T^2 -
MH_T^2}} 
\eeq 
for a di-jet event, where $j_2$ is the less
energetic jet, is very effective for disentangling QCD events, which
typically have $\alpha_T < 0.5$, from events with large MET.
Fig.~\ref{fig:CMSdistributions400} displays various kinematic
distributions for the CMS $\alpha_T$ analysis with fixed $M_{U_1} =
600$~GeV and varying $M_{U_1} - M_{A_1}$.  The distribution exhibiting
the largest variations between the different values of $M_{U_1} -
M_{A_1}$ is that in missing $H_T$ ($MH_T$), whose difference is reflected in
the $MH_T/H_T$ distribution (lower left) and in the $\alpha_T$
variable itself (upper left). On the other hand, the distributions in
$N_{jets}$ and $H_T$ are more stable, though we do observe a slightly
larger average number of jets for larger splitting, due to the
contribution of the jets from the decay.

\subsection{LHC MET Constraints on the Toy Model}

We now apply the LHC constraints to the toy model, so as to gain
insight into the typical limits one may obtain in a generic
extra-dimensional model.  Later we also discuss the differences with a
specific realisation in a complete model including not only a heavy
quark and a stable scalar, but the full set of Kaluza-Klein
excitations, including not only the lowest tier but also higher-mass
recurrences, which have different kinematical behaviours in general.
We note that the signatures of this toy model are similar to a simplified model~\cite{Alwall:2008ag}
used for supersymmetric searches~\cite{SMS},
which contains squarks and the neutralino LSP and produces events with pair production of squarks followed by the 
decay $\tilde{q} \to q \chi_0$. The two models differ in the spins of the new states, which alter
the production cross section and influence the kinematic distributions of the decay products.

As already remarked, in the toy model the production cross section is
largely dominated by the QCD process $p p \to U_1 \bar{U}_1$, which
depends only on the mass of the heavy quark $U_1$.  Subleading
contributions are given by processes like $u u \to U_1 U_1$, which are
mediated by t-channel exchange of the stable boson $A_1$. This is of
electroweak strength, and has been neglected for this analysis.
Although we estimate the experimental sensitivities and bounds with a
single heavy quark, the results can be generalised approximately to
any number of near-degenerate quarks, simply by multiplying the
production cross section by the number of heavy quarks.  The cross
sections for quark-antiquark pair-production processes receive large
corrections from QCD contributions at higher orders. These corrections
must be taken into account when extracting reliable bounds on the
masses of new quarks, and for this purpose we have computed cross
sections at NLO+NNLL using the tool of
Ref.\cite{Cacciari:2011hy,Cacciari_ttbar}, with the results shown in
Table~\ref{sigmastoymodel}.

\begin{figure}
\centering\epsfig{file=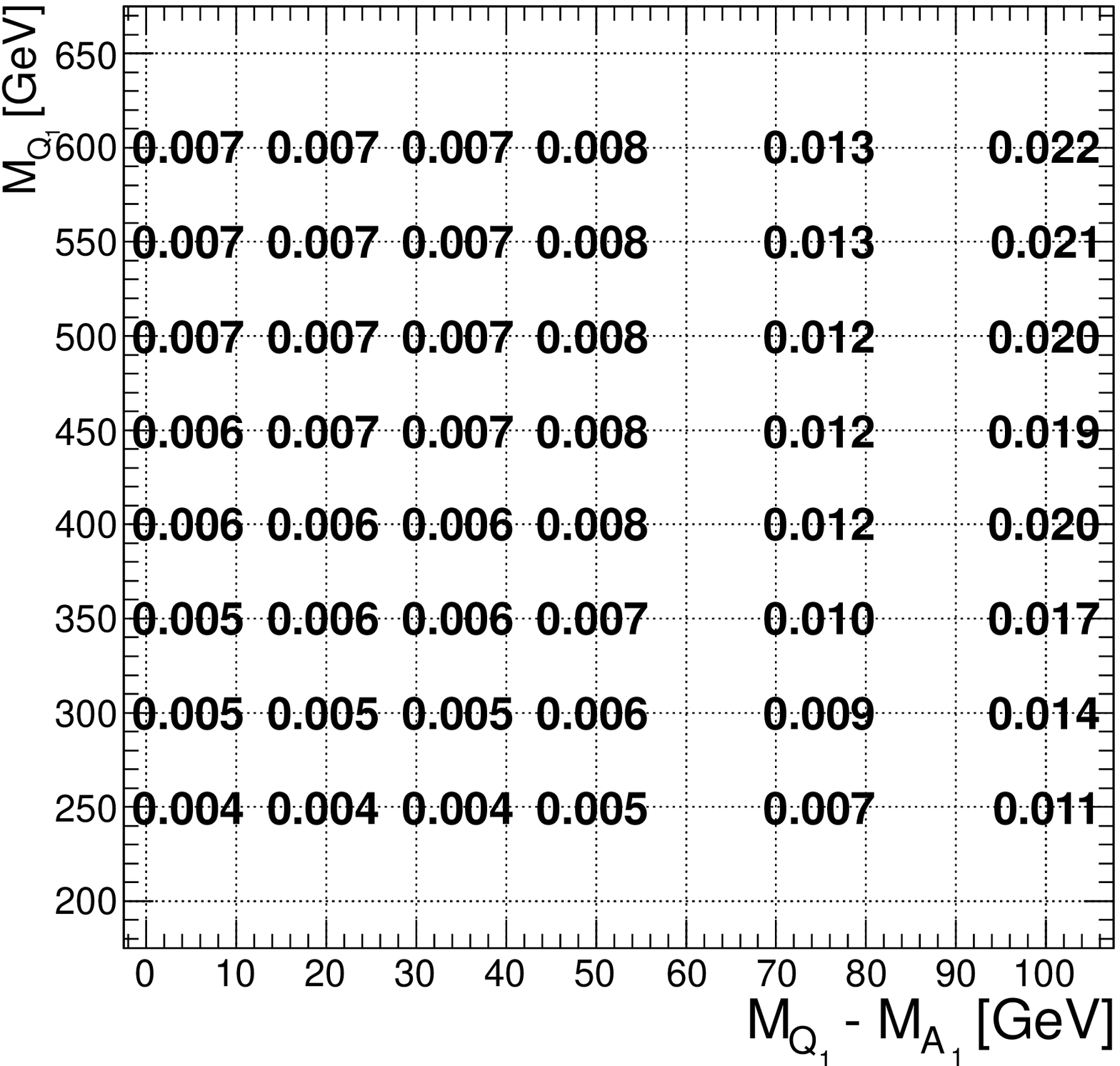,width=.45\textwidth} 
\centering\epsfig{file=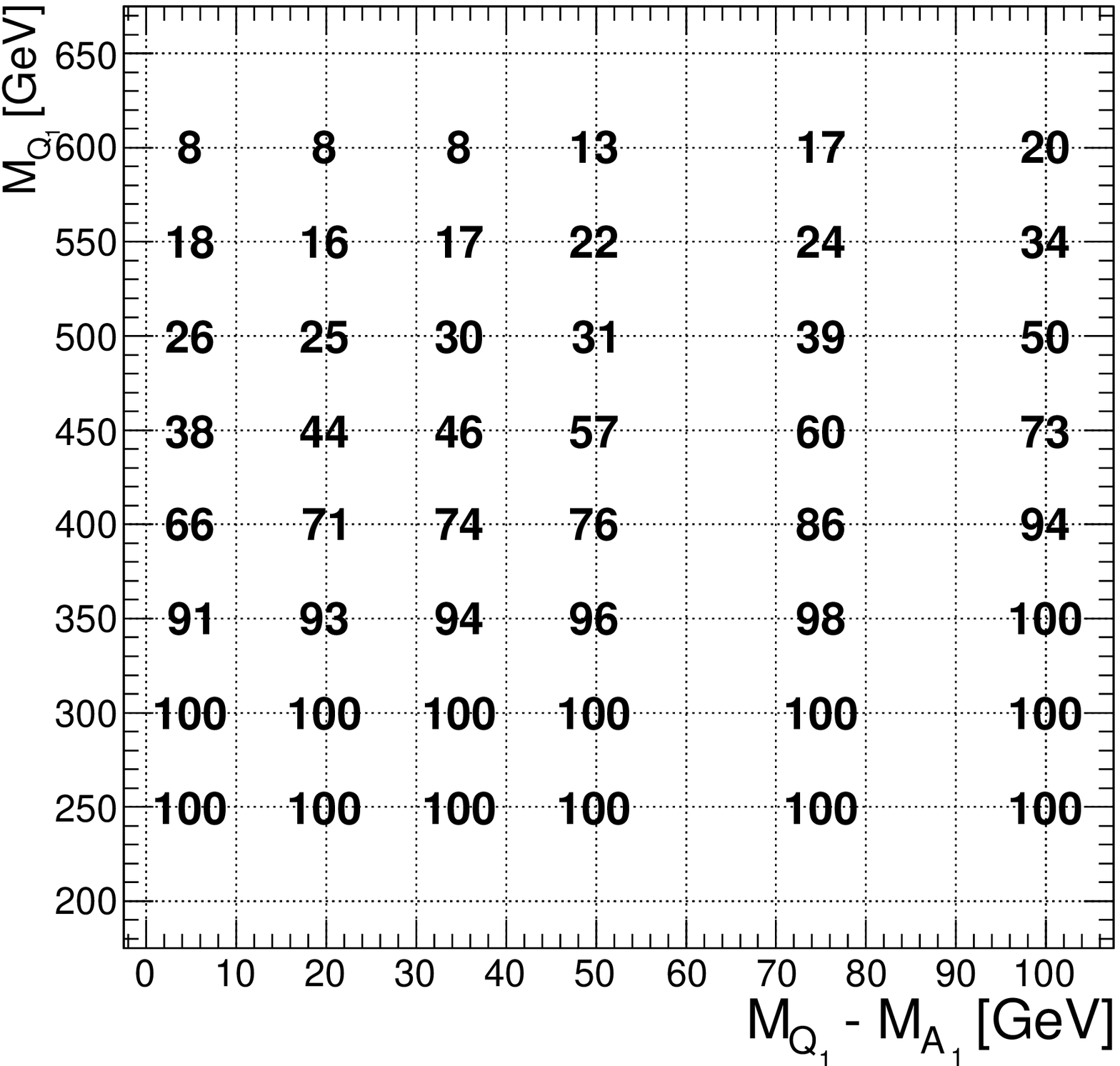,width=.45\textwidth}
\caption{ {\it Efficiency (left) and exclusion confidence level (right) of the CMS $\alpha_T$ analysis for the toy
model, as a function of the heavy quark mass $M_{U_1}$ and the mass
difference $M_{U_1} - M_{A_1}$.}}
\label{fig:efficiency} 
\end{figure}

In the left panel of Fig.~\ref{fig:efficiency} we display the efficiency found in our
{\tt Delphes} simulation of the CMS $\alpha_T$ analysis for the toy
model, as a function of the heavy quark mass $M_{U_1}$ and the mass
difference $M_{U_1} - M_{A_1}$.  We see that the experimental
efficiency for the toy model is of the order $10^{-3}$ to $10^{-2}$,
which is similar to the efficiencies for supersymmetric models with
compressed spectra found in~\cite{LeCompte:2011fh}. Generally
speaking, we see that the efficiency is reduced for smaller mass
differences, essentially because these lead to smaller amounts of MET
for any fixed value of $M_{U_1}$, and is increased for larger values
of $M_{U_1}$, which tend to yield more MET for any fixed value of
$M_{U_1} - M_{A_1}$~\footnote{Typical differences between the masses
of the quark recurrences and the LKP in the realistic model studied
later are $\sim 20$ to 40~GeV, well within the range displayed in
Fig.~\ref{fig:efficiency}: see Fig.\ref{fig:spectra}.}.

The number of signal events expected to be found in the CMS $\alpha_T$
search is the product of the efficiency factor shown in
Fig.~\ref{fig:efficiency} with the cross section shown in
Table~\ref{sigmastoymodel}, for a given integrated
luminosity. However, evaluating the expected significance of any
signal requires a comparison with the background in each bin and its
expected level of fluctuation, which we take from the CMS analysis. In the right panel of
Fig.~\ref{fig:efficiency} we display the confidence level we calculate
using the CL$_s$ statistic for exclusion by the CMS $\alpha_T$
analysis in the toy model, as a function of the heavy quark mass
$M_{U_1}$ and the mass difference $M_{U_1} - M_{A_1}$.  Generally
speaking, the upper limit weakens as the mass difference $M_{U_1} -
M_{A_1}$ decreases, and also as $M_{U_1}$ increases.  The weakening of
the limit as $M_{U_1} - M_{A_1}$ decreases reflects a lowering of the
selection efficiency, as seen in Fig.~\ref{fig:efficiency}, whereas
the increase in the efficiency as $M_{U_1}$ increases does not
counteract the rapid decrease in the cross section.
We see that $M_{U_1} = 300$~GeV is excluded beyond the 99\% CL for any
value of $M_{U_1} - M_{A_1}$, and that $M_{U_1} = 350$~GeV is excluded
at the 95\% CL for mass splittings above and including $M_{U_1} -
M_{A_1} > 50$~GeV.  A similar exercise for the ATLAS experiment
yielded slightly weaker results, mainly due to the lower selection
efficiency for toy model events, so we concentrate on the CMS
$\alpha_T$ search in the following~\footnote{We also found that
monojet searches~\cite{Chatrchyan:2012me} gave significantly weaker levels of exclusion, so we
also do not discuss them further.}.

\section{A More Realistic Model}

In this Section we introduce and analyse a more realistic model with
Universal Extra Dimensions (UED), based on a two-dimensional orbifold
of the Real Projective Plane (RP$^2$)~\cite{Cacciapaglia:2009pa}.
There are two main reasons for the choice of this particular
space. {\it In the first place}, it is the only orbifold in 5 or 6
flat dimensions where an exact Kaluza-Klein K-parity can be defined.
In other realistic models proposed in the
literature~\cite{Appelquist:2000nn,Dobrescu:2004zi}, generic
interactions localised on the fixed or singular points of the orbifold
will in general break the Kaluza-Klein parity of the model explicitly.
Such localised interactions correspond to higher-order operators, so
their absence would require special assumptions about the ultra-violet
completion of the extra-dimensional model.  On the other hand, on the
RP$^2$ the Kaluza-Klein parity is exact, whatever the properties of
the ultra-violet physics.  In addition, the mass splitting between
different states in the same tier, which are generated mainly by loop
corrections, turn out to be rather small compared to other UED models
proposed in the literature~\cite{Cheng:2002iz,Ponton:2005kx}.
Therefore, this scenario is a natural representative of the class of
models we want to explore in this paper.  {\it The second reason} why
we focus on this model comes from the phenomenology of the Dark Matter
candidate: in order for the relic abundance to match the determination
by WMAP and other measurements, the mass scale of the new states is
preferred to lie in the range $700 < m_{KK} < 1000$
GeV~\cite{Arbey:2012ke}.  Higher values of the masses would imply an
overabundance of matter in the Universe, and are therefore excluded by
cosmology~\footnote{We note, however, that by
tuning the mass of the Higgs recurrences, it is possible to 
find a ``funnel'' region, where the annihilation cross section is dominated by resonant s-channel heavy Higgses,
as in the case of supersymmetry,
and the mass scale can be pushed up to $\sim 1.5$ TeV, at the price of fine tuning in the parameter space.}.  
This mass range is ideal for LHC searches, as it offers
production cross sections large enough to be probed with the present
amount of data.  Other models in the literature prefer very different
mass ranges: in the minimal 5-dimensional model, the preferred range
is $1250 < m_{KK} < 1500$ GeV~\cite{Belanger:2010yx}, thus too high to
be accessible at present, whereas for the other 6-dimensional model
based on the chiral square, the preferred range is very low, $180 <
m_{KK} < 200$ GeV~\cite{Dobrescu:2007ec}, thus likely to be excluded
already.

\subsection{A Minimal Scenario with Two Universal Extra Dimensions}

The RP$^2$ orbifold is defined via a discrete symmetry group of a two-dimensional Euclidean
space~\footnote{The RP$^2$ orbifold can also be defined as a sphere with the identification of 
antipodal points. A UED model based on this space has been proposed in~\cite{Dohi:2010vc}, and has very different phenomenology.}
describing the two extra space co-ordinates ${x_5, x_6}$. It consists of a reflection, $r$, and a glide, $g$, whose actions on the co-ordinates are:
\bea
r(\{x_5,x_6\}) &= & \{- x_5, - x_6\}\,, \\ 
g(\{x_5,x_6\}) & = & \{x_5 + \pi R_5, - x_6 + \pi R_6\}\,.
\eea
We note that $g' = g*r$ defines a second glide that flips the sign of $x_5$, 
and that, squaring the two glides, one obtains translation symmetries along the two co-ordinate directions: 
\bea
g*g(\{x_5,x_6\}) & = & \{x_5 + 2 \pi R_5, x_6\}\,, \\ 
g'*g'(\{x_5,x_6\}) & = & \{x_5, x_6 + 2\pi R_6\}\,.
\eea
The fundamental space of RP$^2$ can therefore be thought of as a subspace of an orthogonal torus with radii $R_5$ and $R_6$.

The spectrum of the model is therefore the same as a torus, up to a projection defined by the orbifold symmetries: 
each tier of resonances can be labelled by two integer numbers $(n,m)$ that represent units of the quantised momenta 
along the two extra directions. The bulk mass is therefore given by
\bea
m_{(n,m)}^2 = \frac{n^2}{R_5^2} + \frac{m^2}{R_6^2} \, ,
\eea
where the two inverse radii define the mass scales in the model.
The field content of each tier of recurrences depends on the field content and on the parity assignment on 
each six-dimensional field.

In the following, we focus on a UED model in which a bulk field is associated to each Standard
Model (SM) field~\cite{Cacciapaglia:2009pa}. Note that chiral fermions are independent fields in the 
SM, therefore two 6-dimensional fields are associated to each SM fermion. 
Parities are assigned to each field in such a way that the zero-mode spectrum coincides with that
in the SM. The conserved symmetry that is responsible for the stability of the Dark Matter candidate
can be defined in terms of a translation
\bea
P_K  (\{x_5,x_6\}) = \{ x_5 + \pi R_5, x_6 + \pi R_6 \}\,,
\eea
and the tiers have parity $(-1)^{n+m}$ under this symmetry.
This is the only exact symmetry that is respected by all interactions in the model~\cite{Cacciapaglia:2012dy}. 
The degeneracy within the tiers of recurrences with the same values of $(n, m)$
is lifted by loop corrections and by the contribution of the Higgs VEV.

In the following we are mainly interested in the lightest tiers: the odd tiers $(1,0)$ and $(0,1)$, and the even tiers $(2,0)$ and $(0,2)$.
Each odd tier contains a gauge scalar boson corresponding to each SM gauge vector boson,
and a vector-like fermion for each chiral Standard Model fermion.
To simplify the notation, we label the odd states with a subscript ``$1$'': the complete field content of the odd tiers is therefore
\bea
\mbox{gauge scalars} & \Rightarrow &  A_1,~Z_1,~W_1^\pm,~G_1\,; \nonumber\\
\mbox{fermions}  &  \Rightarrow & q_{1S},~q_{1D},~l^\pm_{1L},~l^\pm_{1R},~\nu^l_1\,, \nonumber
\eea
where $q = u, d, s, c, b, t$ labels the quarks, $l = e, \mu, \tau$ the charged leptons.
The odd states in these tiers can only decay to the lightest state in the tier, which is the Dark Matter candidate $A_1$, plus a SM particle.
The even tiers contain a vector gauge boson corresponding to each Standard Model gauge boson, 
a vector-like fermion for each Standard Model chiral fermion and a massive scalar Higgs doublet.
We denote even states with a subscript ``$2$'':
\bea
\mbox{gauge vectors} & \Rightarrow &  A^\mu_2,~Z^\mu_2,~W_2^{\mu \pm},~G^\mu_2\,; \nonumber\\
\mbox{fermions}  &  \Rightarrow & q_{2S},~q_{2D},~l^\pm_{2L},~l^\pm_{2R},~\nu^l_2\,; \nonumber\\
\mbox{Higgs scalars}  &  \Rightarrow & H_{2},~S^0_{2},~S^\pm_{2}\,. \nonumber
\eea
Here the $S^{0,\pm}_2$ are the recurrences of the Goldstone bosons of the SM Higgs, i.e., 
the charged component of the doublet and the imaginary part of the neutral field (pseudoscalar).
Depending on the mass corrections, a state in an even tier may decay into a pair of odd $(1)$ states, 
into an even state $(2)$ plus a Standard Model particle, or into a pair of Standard Model particles.
The latter would arise from a loop-induced coupling~\cite{Cacciapaglia:2011hx}, but the rate is comparable to the other two processes, 
which are suppressed by the phase space available after the loop-induced mass corrections.
The MET signal comes typically from the production of odd states either directly or via the decays of an even state~\footnote{We 
note that there is another even tier that is potentially lighter that the $(2,0)$ and $(0,2)$ one, namely the $(1,1)$ tier.
However, it can decay into a pair of Standard Model states only via localised higher-order operators, 
whilst the tree-level process $(1,1) \to (1,0) + (0,1)$ is below threshold.
We will not consider this tier any further, because it does not lead to MET signals. 
Some aspects of its phenomenology have been studied in~\cite{Cacciapaglia:2011kz}.}.

The spectra of the four tiers we are interested in depend on two free parameters, namely the two radii $R_5$ and $R_6$.
The loop corrections are also logarithmically sensitive to a cut-off, i.e., the scale where the 6-dimensional
theory is no longer under perturbative control. In the following, we fix this scale to be $10$ times the mass scale of the
Kaluza-Klein states, which is the largest value allowed by naive dimensional analysis.
Lowering the value of the cut-off would decrease the mass splitting, but also increase the sensitivity of 
direct Dark Matter detection experiments to the level that values of the 
cut-off below $4 m_{KK}$ are already excluded~\cite{Arbey:2012ke}.
The spectrum also depends on the higher-order operators, localised on the two singular points, 
which act as counter-terms for logarithmic divergences in the loop corrections. In the spirit of effective field theories, 
we assume that the contributions of the higher-order operators are negligible compared to the loop corrections.
When writing the loop correction to the mass as $m = m_{(n,m)} (1+\delta)$, it
should be noted that $\delta$ has a mild dependence on the ratio $R_5/R_6$. This is due to the fact that the main 
contribution to the loop correction is proportional to the Kaluza-Klein bulk mass of the state~\cite{Cacciapaglia:2012dy}.
The localised operators will modify the spectrum also by introducing mixing between the two odd and the two even tiers.
However, while the even tiers can also mix via loop contributions, the odd ones only mix via contributions of the 
ultraviolet physics. In our phenomenological analysis, therefore, we need only consider mixing in the even sector, 
which can be rather large if the values of the two radii are close.
This fact allows us to distinguish two separate regimes: one where the radii are almost degenerate, 
their values being similar within 10\%, and one where there is a hierarchy between the two radii.
The main difference between the two limits is in the preferred mass range~\cite{Arbey:2012ke}: 
in the degenerate limit, the mass scale tends to be smaller and is already excluded for all values of the cut-off by
direct Dark Matter detection experiments.
On the other hand, in the hierarchical case, only one pair of sets of states associated with the larger radius, 
say $(1,0)$ and $(2,0)$, are relevant, and the preferred mass range is $700 < m_{KK} < 1000$ GeV, 
with values of the cut-off below 4 excluded by direct Dark Matter detection experiments.

\begin{figure}
\epsfig{file=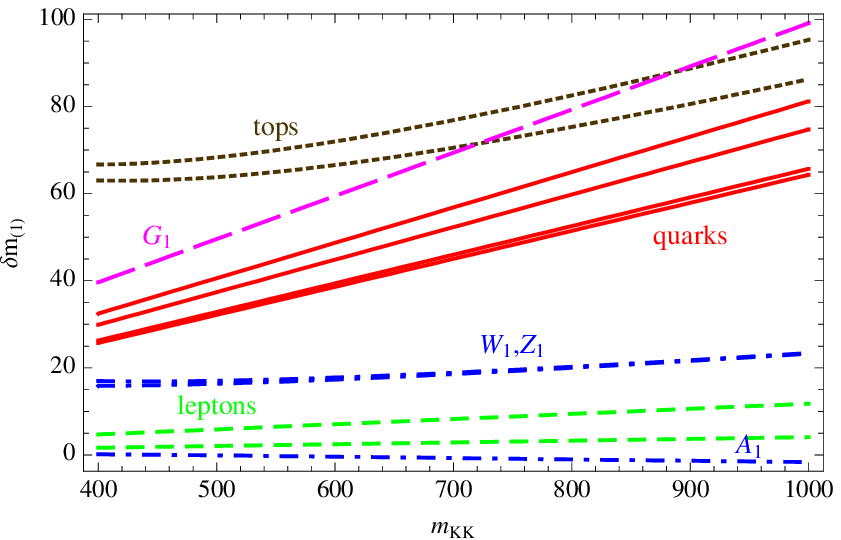,width=.5\textwidth}
\epsfig{file=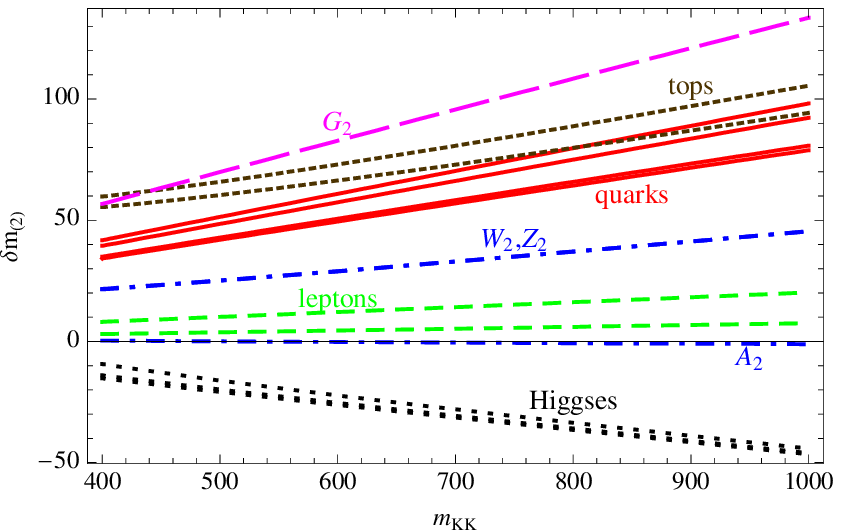,width=.5\textwidth}
\caption{\it Mass correction for the lightest odd (1) and even (2) tiers of recurrences in the hierarchical limit. We show the correction with respect to the bulk mass, respectively $m_{KK}$ and $2 m_{KK}$. The various colours label recurrences of: the electroweak gauge bosons in blue ($A$, $W^\pm$ and $Z$, from bottom to top), leptons in green (singlets and doublets), light quarks in red (up and down singlets, doublet ant bottom-doublet), top quarks in violet (singlet/doublet), gluon in magenta and higgses in black ($S^\pm_{2}$, $S_{2}$ and $H_{2}$, where $S$ are recurrences of the Goldstone bosons).}
\label{fig:spectra}
\end{figure}

In the following, we only consider the hierarchical limit:
in this case, the phenomenology is dominated by the lightest tiers, which get their masses from the larger radius.
The results therefore only depend on one parameter, $m_{KK} = min(1/R_5, 1/R_6)$, which sets the overall mass scale.
In Figure~\ref{fig:spectra} we show the mass corrections $\delta m_{(1)} = m_{(1)} - m_{KK}$ 
for the lightest odd tier as a function of $m_{KK}$ at one loop.
The figure shows that the mass difference between each KK recurrence and the lightest one over the 
relevant mass range amounts to a few tens of GeV for coloured states (quarks and gluon recurrences), 
i.e., within the range studied previously in the toy model, and few GeV for leptons.
In the right panel of the same figure we show the same mass corrections for the even tier, $\delta m_{(2)} = m_{(2)} - 2 m_{KK}$.

\subsection{Simulation of the RP$^2$ Scenario}

The MET signals we study in this work are generated by the decays of odd states into the dark matter candidate $A_1$.
In the following, we consider two classes of signals ($X_{1,2}$ labels any particle in the odd or 
even tiers)~\footnote{Another potentially interesting class is $p p \to X_2 \to X_1 X_1$ and $p p \to X_2 \to x_{\rm SM} X_2$, 
where the resonant even state is a gauge vector ($G_{2}$, $W^\pm_{2}$, $Z_{2}$ or $A_{2}$) and its production takes place via a 
loop-induced vertex. We found that the matching procedure used in this work has difficulties with the latter process, 
where $x_{\rm SM}$ is a jet, therefore including this process would entail calculations beyond the scope of this work. 
To be conservative, we did not consider single resonant production.}
\begin{enumerate}
\item[1)] $p p \to X_1 X_1$, which is dominated by the production of quarks, and can
therefore be compared with the toy model in the previous section;
\item[2)] $p p \to X_2 X_2$, where one of the even states decays into a pair of odd states 
and the other decays into energetic SM particles (jets, tops or leptons) thus providing the visible energy for the analysis.
\end{enumerate}

The processes in class 1) give the largest production rates, due to the smaller masses, and they have similar kinematics as the toy model.
The processes in class 2) are particularly interesting, as they are absent in minimal supersymmetric and Little Higgs models, 
where all (or most of) the new states are odd under the dark matter parity. The complete RP$^2$ model, including the loop-induced 
vertices, has been implemented in {\tt MadGraph5} with the help of the {\tt FeynRules}~\cite{Christensen:2008py} package, 
and tested by comparing branching ratios and cross sections with the {\tt CalcHEP 3.0}~\cite{Pukhov:1999gg} implementation.
In the implementation we included the full one-loop results for the masses of both even and odd recurrences,
which depend only very mildly on the ratio of the two radii~\cite{Cacciapaglia:2012dy}, as noted previously.
The mass splittings between various states in the odd $(1)$ and even $(2)$ tiers are shown in Figure~\ref{fig:spectra}.
Since the mass differences between coloured and electroweak states are typically a few tens of GeV, 
production at the LHC is dominated by coloured states.
The decay chains we are interested in are therefore initiated by heavy quarks and gluon recurrences.
These chains can be very complicated, because they can involve the lighter $W$ and $Z$ recurrences and the leptons.
For the even tiers, decays into a pair of resonant SM particles can also occur with a significant rate.
Quarks, for instance, always decay into a SM quark plus an electroweak gauge boson in the same tier: 
this is true for both odd and even recurrences.
The subsequent decays of the gauge bosons will involve further jets and, in some cases, leptons.

\begin{figure}
$$m_{KK}=400~GeV$$
\begin{minipage}{.5\textwidth}
\epsfig{file=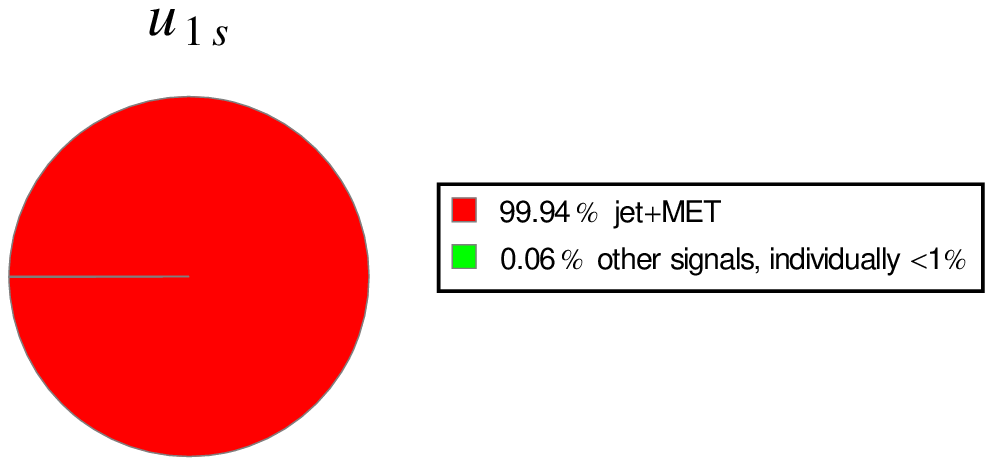,width=\textwidth}
\end{minipage}
\begin{minipage}{.5\textwidth}
\epsfig{file=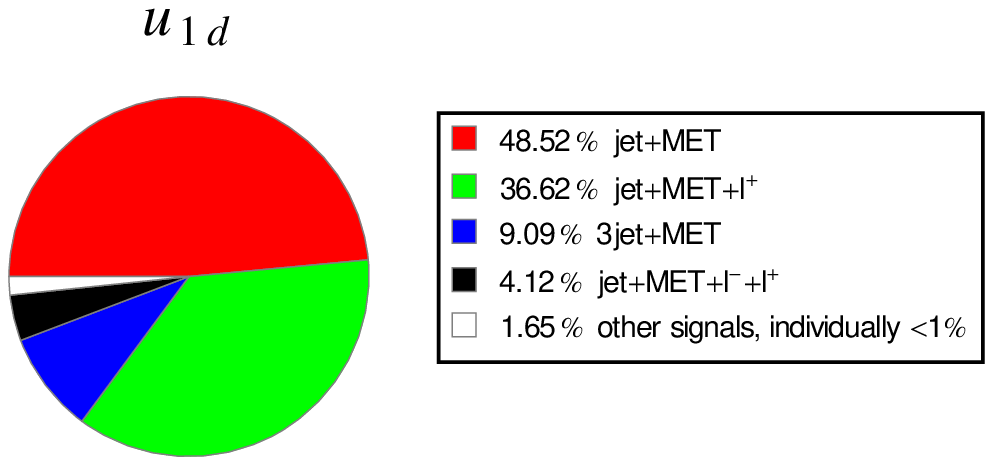,width=\textwidth}
\end{minipage}
\begin{minipage}{.5\textwidth}
\epsfig{file=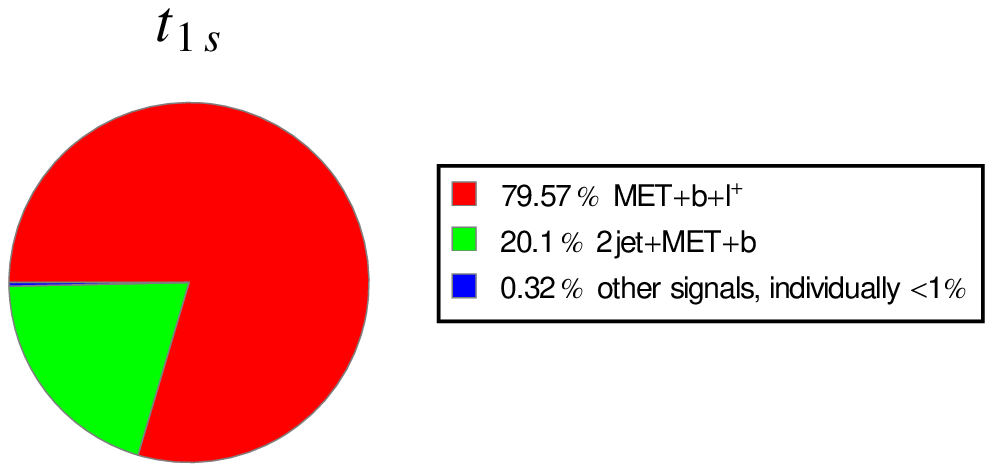,width=\textwidth}
\end{minipage}
\begin{minipage}{.5\textwidth}
\epsfig{file=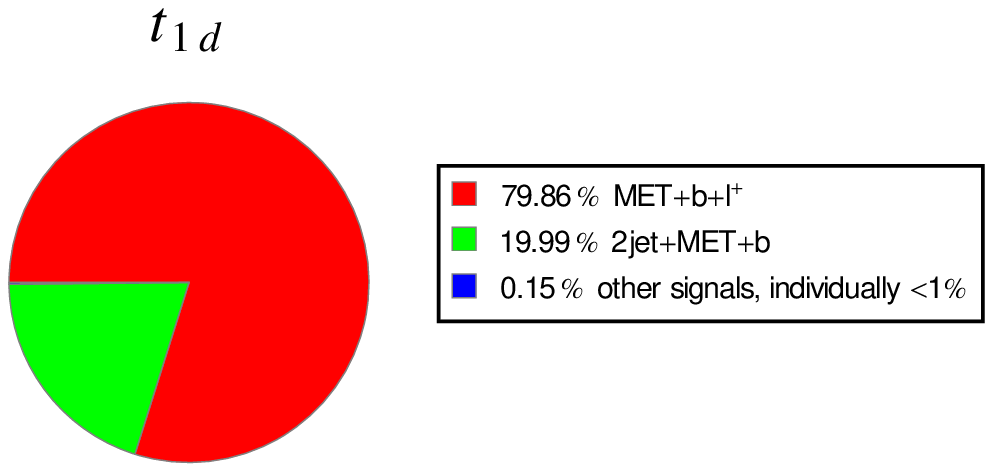,width=\textwidth}
\end{minipage}
\caption{\it Branching ratios for quarks of the (1,0) tier decaying to final states containing jets, leptons and missing energy. 
Only partners of the up and top quark are shown: the branching ratios of other quark
recurrences are similar to those of the up recurrences.}
\label{fig:BR1}
\end{figure}

The possible final states of the decay chains of the odd quarks, grouped according to the numbers of jets and leptons 
(electrons, muons and leptonically decaying taus), are pictured in Figure~\ref{fig:BR1} 
together with their relative rates, for a benchmark value $m_{KK} = 400$ GeV.
The relative rates change only slightly for larger masses.
These plots show that the singlet recurrences $q_{1S}$ decay predominantly into a single jet plus a Dark Matter candidate,
and so correspond directly to the toy model.
On the other hand, doublet recurrences $q_{1D}$ yield significant rates for final states with leptons or multiple jets,
so they reflect physics not included in the toy model.
The top recurrences preferably decay into final states with leptons, 
however their production cross section is much smaller compared to the recurrences of light quarks, 
thus their impact in the phenomenology is limited.

\begin{figure}
$$m_{KK}=400~GeV$$
\begin{minipage}{.5\textwidth}
\epsfig{file=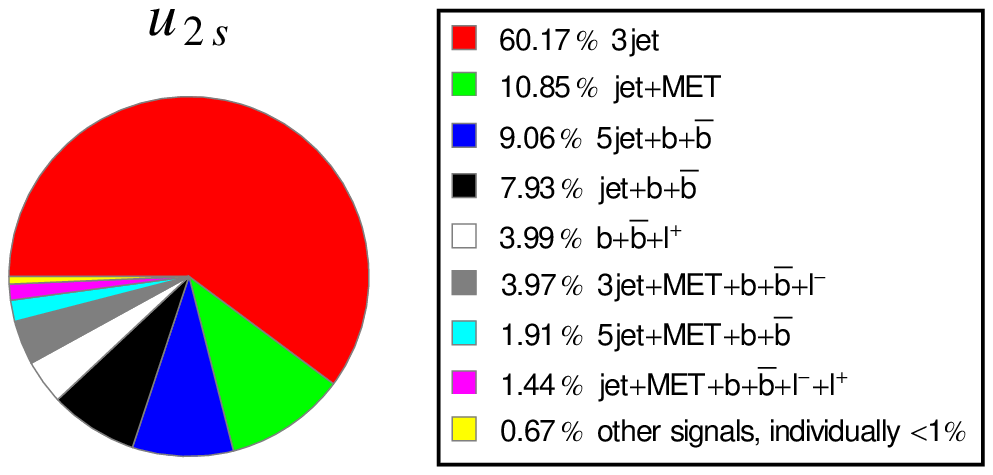,width=\textwidth}
\end{minipage}
\begin{minipage}{.5\textwidth}
\epsfig{file=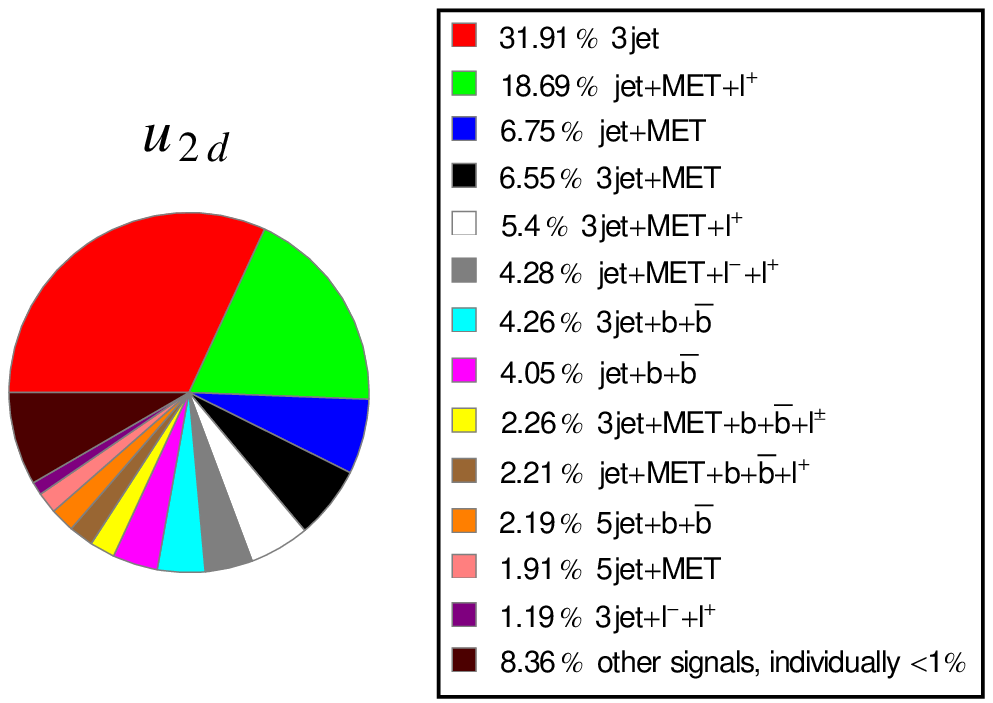,width=\textwidth}
\end{minipage}
\begin{minipage}{.5\textwidth}
\epsfig{file=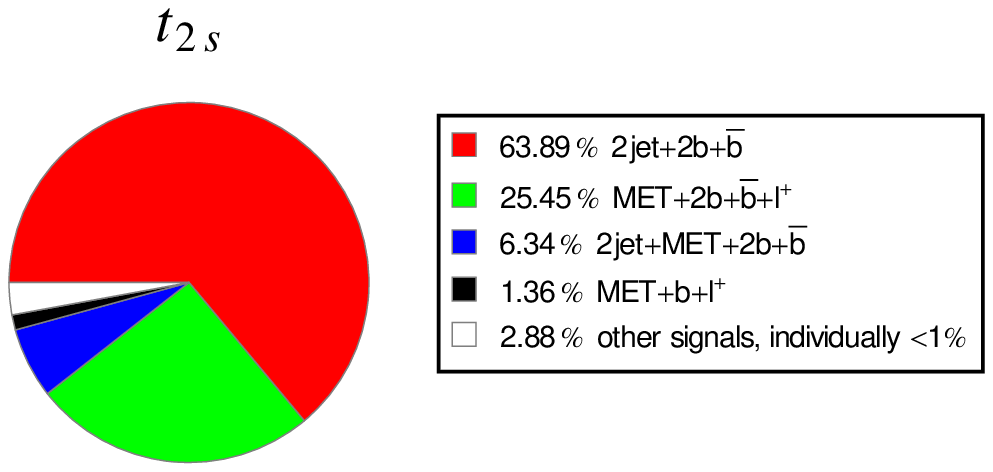,width=\textwidth}
\end{minipage}
\begin{minipage}{.5\textwidth}
\epsfig{file=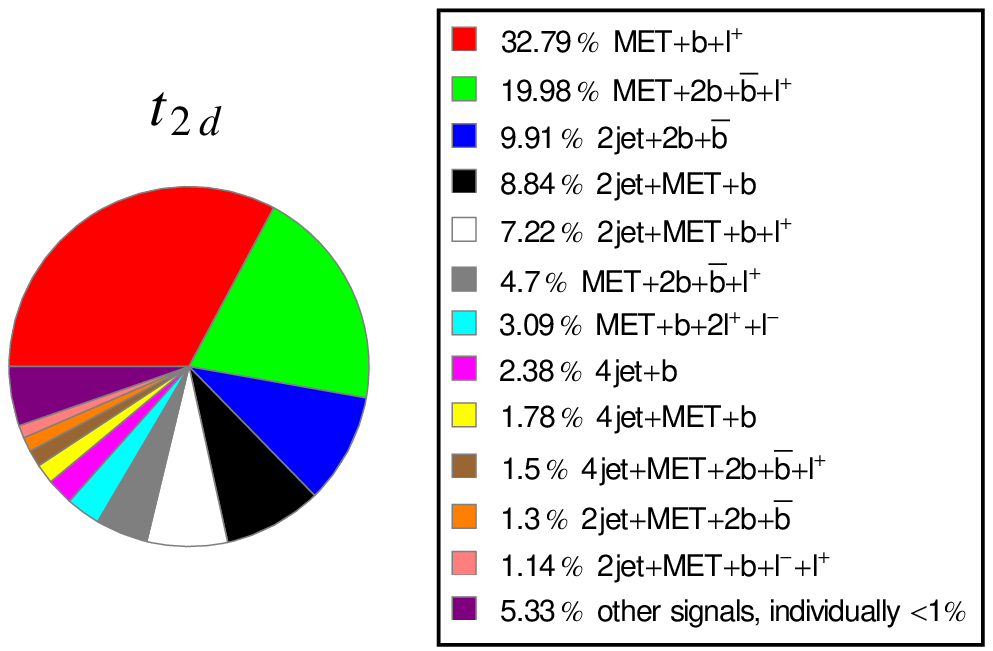,width=\textwidth}
\end{minipage}
\caption{\it Branching ratios for quarks of the (2,0) tier decaying to final states containing jets, leptons and missing energy.
Only partners of the up and top quark are shown: the branching ratios of other quark
recurrences are similar to those of the up recurrences.}
\label{fig:BR2}
\end{figure}

\begin{figure}
$$m_{KK}=400~GeV$$
\begin{minipage}{.5\textwidth}
\epsfig{file=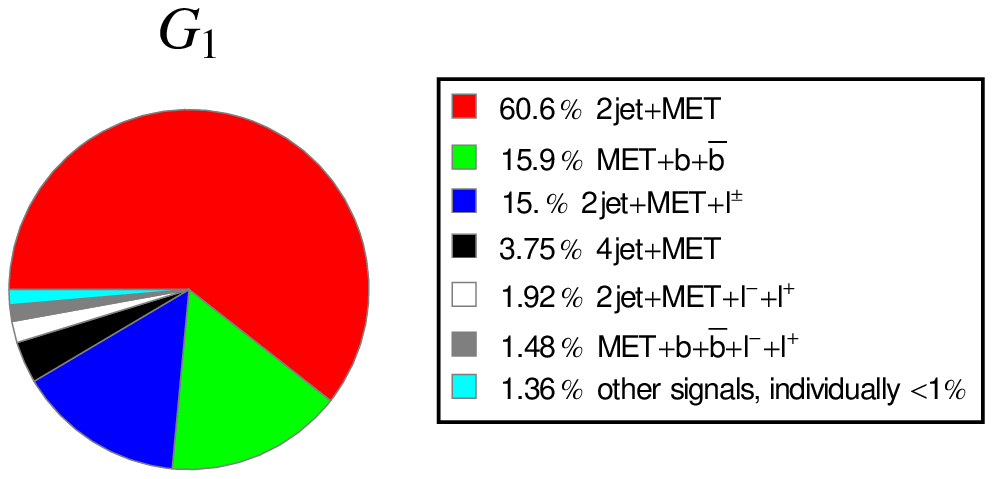,width=\textwidth}
\end{minipage}
\begin{minipage}{.5\textwidth}
\epsfig{file=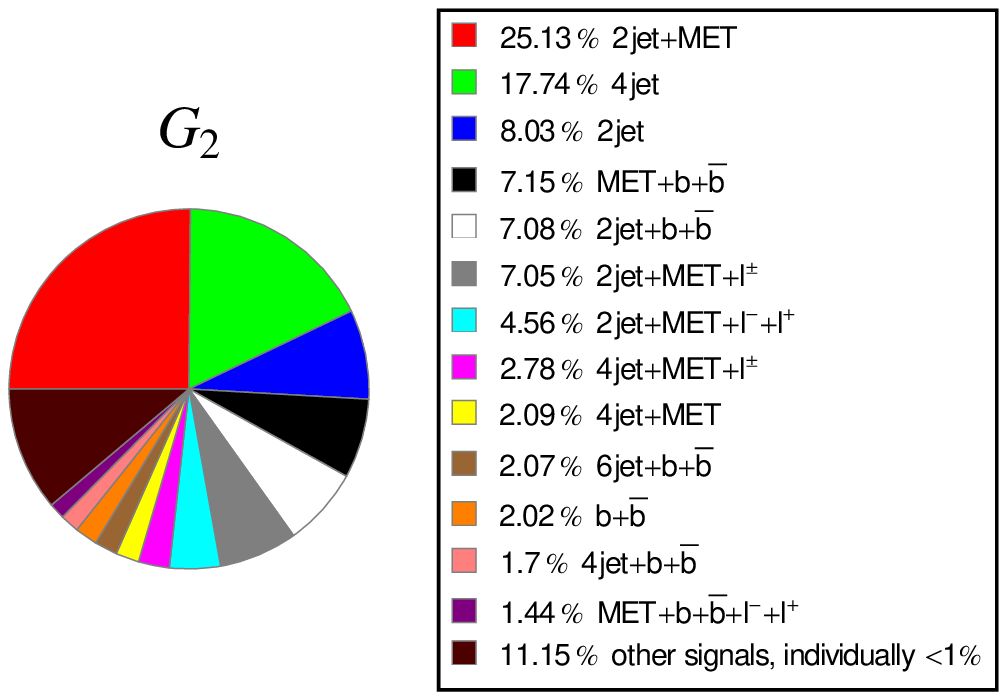,width=\textwidth}
\end{minipage}
\caption{\it Branching ratios for gluons of the (1,0) and (2,0) tiers.}
\label{fig:BR3}
\end{figure}

Similar plots for the quark even recurrences are shown in Figure~\ref{fig:BR2}: here we note large rates in final states without 
MET, due to the resonant decays of the heavy gauge bosons in the chains.
We note also that here the MET is due to a pair of Dark Matter particles, so the kinematic distributions generated by the 
decays of these particles are very different from the toy model, as we see later.
The branching ratios for the odd and even gluon recurrences are shown in Figure~\ref{fig:BR3}.

In producing signal events, we have considered only the most relevant production processes of heavy states,
namely those due to QCD interactions. Their production channels can be grouped in three sets:
\begin{eqnarray}
G_i G_i \quad G_i Q_i \quad Q_i Q_i
\end{eqnarray}
where $i=1,2$ and $Q_i$ stands for both quarks and anti-quarks.
In addition to the usual QCD interactions, the model also contains processes mediated by the heavy gluon recurrence: 
for instance, significant contributions come from the process $u u \to U_i U_i$, mediated by t-channel exchange of the gluon $G_i$.
Such contributions are included in our simulation.
We checked that the contribution of electroweak processes is negligible, amounting to a percent level of the QCD cross section at most.

\begin{table}[tb]
\footnotesize
\centering\begin{tabular}{|c|cccc|cccc|c|}
\hline
                    & \multicolumn{4}{c|}{Odd tier} & \multicolumn{4}{c|}{Even tier} & \\
$M_{KK} (GeV)$   &  $Q_1Q_1$ & $Q_1 \overline Q_1$ &  $G_1Q_1$  & $G_1G_1$ & $Q_2Q_2$            &  $Q_2 \overline Q_2$ &  $G_2Q_2$            & $G_2G_2$  & total\\
\hline
400  &  1,630 & 7,440 & 4,780 & 418 & 718 & 159 & 476 & 43 & 15,700 \\
600 &  221 & 531 & 327 & 18 & 25.8 & 2.6 & 7.2 & 0.4 & 1,130 \\
700 & 99 & 179 & 119 & 5.7 & 5.2 & 0.36 & 1.08 & 0.05 & 409 \\
\hline
\end{tabular}
\caption{\it Production cross sections in fb at $7$~TeV for the 8 classes discussed in the text, 
which are distinguished by the pairs of odd and even states produced.}
\label{tab:crosssections}
\end{table}

The framework for the simulation and jet matching is analogous to that used for the toy model. 
The main difference arises from the rich structure of chain decays of RP$^2$ states,
which leads to a great variety of different final states. A simulation of production and decay of 
heavy RP$^2$ states within {\tt Madgraph} would be computationally extremely demanding. Therefore,
we have used an external tool to decay heavy particles in the events, taking into account the helicity amplitudes,
namely the software {\tt BRIDGE} \cite{Meade:2007js}, which decays particles in event files into every possible final state, 
taking into account spin correlations. This procedure is, however, an approximation because spin correlation 
effects between the {\tt Madgraph} output and the {\tt BRIDGE} input cannot be fully taken into account, 
but kinematical properties of final states are reproduced with a reasonable accuracy. 
In any case, the detector simulation smears physical observables and, moreover, 
we do not search for asymmetries or polarisation effects in the final states. 
To analyse a specific final state it is therefore enough to generate a large sample of events 
and select the subset of events only in the final phase of the analysis.

In making our simulation we grouped similar production processes into 8 classes: 
$G_1 G_1$ and $G_2 G_2$ comprising pair production of gluons, 
$G_1 Q_1$ and $G_2 Q_2$ comprising the production of a gluon with a quark or anti-quark, 
$Q_1 Q_1$ and $Q_2 Q_2$ comprising pair production of two quarks~\footnote{The production of two antiquarks is negligible.}, 
and finally $Q_1 \bar{Q}_1$ and $Q_2 \bar{Q}_2$ comprising the production of a quark and antiquark.
In all classes, we include all flavours of the heavy and SM quarks and, as in the toy model, 
processes with 1 and 2 additional jets are included.
The matching procedure has been optimised for each class and each mass choice independently, 
following the procedure described in Section~\ref{sec:simulation}.
We studied three benchmark points, with masses $m_{KK} = 400$, $600$ and $700$ GeV.
The cross sections in each of the individual classes have been computed at leading order using our {\tt Madgraph} implementation
and are shown in Table~\ref{tab:crosssections}.
Unlike in the toy model, it was not possible to reliably estimate the QCD corrections to the cross section, 
due to the presence of many diagrams contributing to each production class, and to the presence of heavy gluons.
Therefore, we conservatively used LO cross sections in this study.
For each production class and mass point we generated 300,000 events, with a large majority passing the 
matching procedure with {\tt PYTHIA}.

\subsection{MET Searches with Leptons}

For the analysis, we used the same validated {\tt Delphes} implementation  of the CMS and ATLAS detectors
that we used to analyse the toy model. However, because of the presence of leptons in the final states, 
which were absent in the toy model, we now include, together with the MET analysis detailed in Section~\ref{sec:analysis},
typical supersymmetry-motivated searches based on leptons. Thus, following the philosophy of the previous section, we focused on the searches published by the CMS collaboration: together with the CMS $\alpha_T$ analysis~\cite{Chatrchyan:2011zy}, 
in this section we consider single-lepton~\cite{CMSLpsearch}, opposite-sign~\cite{CMSOSsearch} 
and same-sign dilepton~\cite{CMSSSsearch} searches.
Similar searches have been performed by ATLAS, and their inclusion would not change the conclusion of our work.
A study of the reach of the ATLAS multi-lepton searches for supersymmetric compressed spectra can be found in Ref.~\cite{Rolbiecki:2012gn}.

The single-lepton search relies on events with a single high-$p_T$ lepton together with jets and MET.
Three different methods are used to discriminate the signal from the background, 
and the one we implement here is based on the ``lepton projection'' method that uses a variable $L_p$~\cite{CMSLpsearch}, 
which measures the component of the lepton transverse momentum that is parallel to that of the reconstructed $W$ it originates from.
For SM events, where the lepton comes from the decay of a $W$ boson via a left-handed coupling, 
the lepton momentum tends to be aligned in the direction of motion of the $W$, resulting in a large value of $L_p$.
Therefore, a signal region is defined for $L_p < 0.15$, while $L_p > 0.3$ is used as a control region.
Additional cuts on $H_T$ and on the scalar sum of the transverse momentum of the lepton and transverse missing momentum, 
$S_T^{lept}$, are also implemented (typically, $H_T > 500$ GeV).
In the opposite-sign (OS) dilepton search~\cite{CMSOSsearch}, events with a pair of isolated leptons are selected, 
rejecting regions close to the $Z$ peak in invariant mass.
To suppress the $t \bar{t}$ background, 4 signal regions are defined with cuts on the missing transverse energy and $H_T$: 
high-$E_T^{miss}$ ($E_T^{miss} > 275$ GeV, $H_T > 300$ GeV), high-$H_T$ ($E_T^{miss} > 200$ GeV, $H_T > 600$ GeV), 
tight signal region ($E_T^{miss} > 275$ GeV, $H_T > 600$ GeV) and low-$H_T$ ($E_T^{miss} > 275$ GeV, $125< H_T < 300$ GeV).
Finally, in the same-sign (SS) search~\cite{CMSSSsearch}, the presence of two isolated dileptons with the same charge is required, 
where the isolation is established by requiring that the scalar sum of the track momenta and calorimeter energies in a 
cone $\Delta R < 0.3$ around the lepton track has to be less than 15\% of the lepton momentum.
Five signal regions are defined with various cuts on $H_T$ and $E_T^{miss}$: 
Region 1 ($H_T > 80$ GeV, $E_T^{miss} > 120$ GeV),  Region 2 ($H_T > 200$ GeV, $E_T^{miss} > 120$ GeV),  
Region 3 ($H_T > 450$ GeV, $E_T^{miss} > 50$ GeV),  Region 4 ($H_T > 450$ GeV, $E_T^{miss} > 120$ GeV), 
and Region 5 ($H_T > 450$ GeV, $E_T^{miss} > 0$ GeV). 

\begin{table}[tb]
\footnotesize
\centering\begin{tabular}{|c|ccc|ccc|ccc|ccc|}
\hline
                    & \multicolumn{3}{c|}{$\alpha_T$} & \multicolumn{3}{c|}{$L_P$} & \multicolumn{3}{c|}{$OS$} & \multicolumn{3}{c|}{$SS$} \\
$M_{KK} (GeV)$      & $400$ & $600$ & $700$ & $400$ & $600$ & $700$ & $400$ & $600$ & $700$ & $400$ & $600$ & $700$ \\
\hline
$Q_2Q_2$            & 242.84& 6.13  & 1.21  & 73.65 & 3.39  & 0.86  & 11.17 & 0.55  & 0.15  & 6.01  & 0.28  & 0.09  \\
$Q_2 \overline Q_2$ & 41.66 & 0.56  & 0.06  & 14.66 & 0.37  & 0.05  & 2.38  & 0.07  & 0.01  & 1.15  & 0.03  & 0.00  \\
$G_2G_2$            & 10.34 & 0.08  & 0.01  & 2.74  & 0.02  & 0.00  & 0.4   & 0.00  & 0.00  & 0.19  & 0.00  & 0.00  \\
$G_2Q_2$            & 125.33& 1.57  & 0.25  & 36.74 & 0.73  & 0.15  & 5.07  & 0.11  & 0.03  & 2.39  & 0.06  & 0.01  \\
$Q_1Q_1$            & 64.68 & 8.51  & 3.3   & 2.17  & 0.26  & 0.11  & 0.19  & 0.04  & 0.04  & 0.19  & 0.02  & 0.01  \\
$Q_1 \overline Q_1$ & 342.86& 24.06 & 7.29  & 12.53 & 0.86  & 0.24  & 2.6   & 0.29  & 0.16  & 0.5   & 0.04  & 0.03  \\
$G_1G_1$            & 28.38 & 1.7   & 0.61  & 0.65  & 0.05  & 0.02  & 0.02  & 0.00  & 0.00  & 0.01  & 0.00  & 0.00  \\
$G_1Q_1$            & 229.58& 20.78 & 8.51  & 6.29  & 0.63  & 0.27  & 0.4   & 0.08  & 0.06  & 0.24  & 0.03  & 0.02  \\
\hline
Total               &1085.67& 63.39 & 21.24 & 149.43& 6.31  & 1.7   & 22.23 & 1.14  & 0.45  & 10.68 & 0.46  & 0.16 \\
\hline 
\end{tabular}
\caption{\it Numbers of events for each subprocess that pass the signal selections of the respective searches in 5/fb at 7~TeV.}
\label{tab:searchoverview}
\end{table}

\begin{figure}
\begin{center}
\begin{minipage}{.8\textwidth}
\epsfig{file=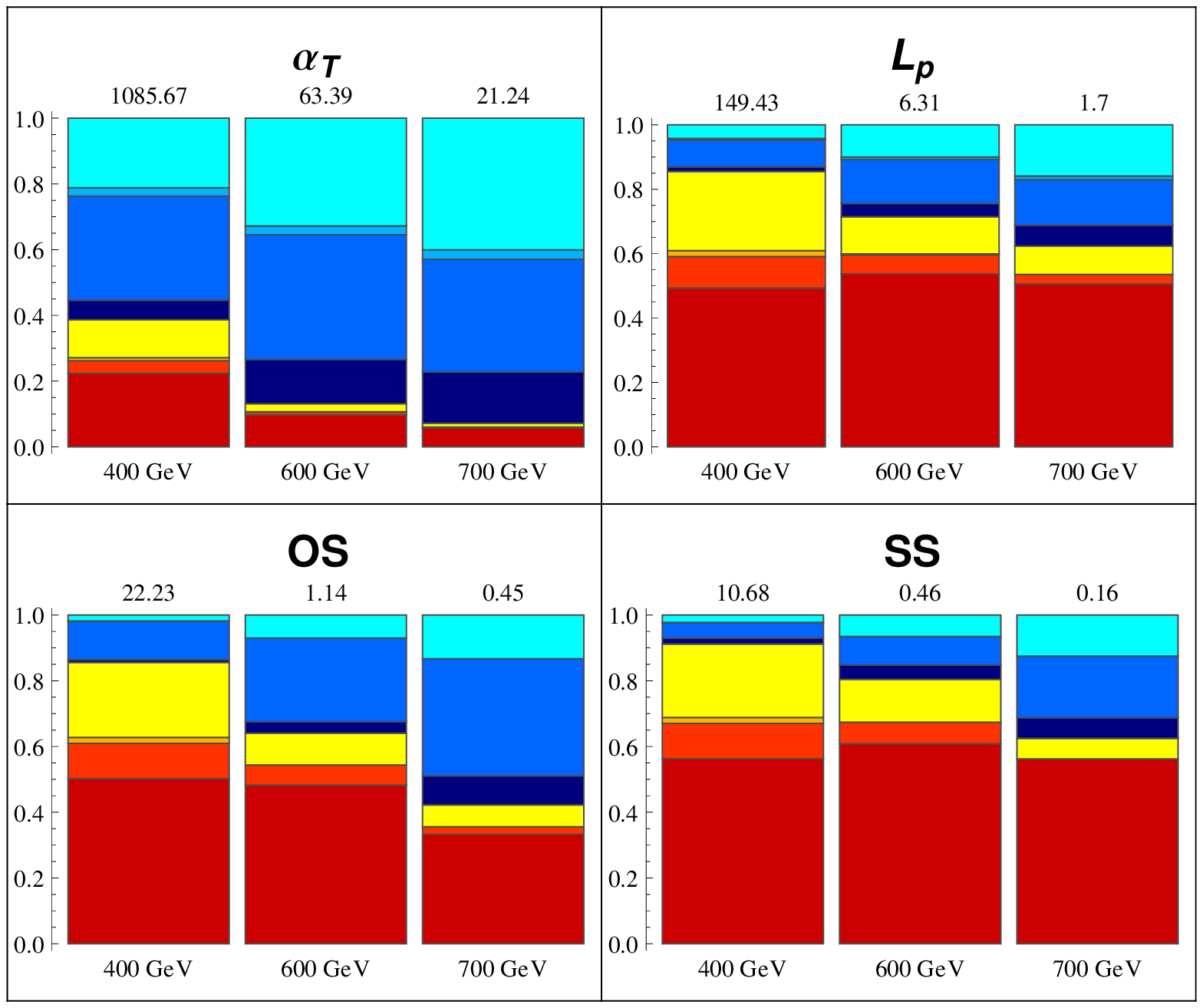,width=\textwidth}
\end{minipage}
\begin{minipage}{.1\textwidth}
\epsfig{file=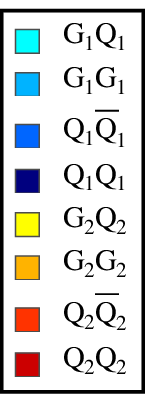,width=\textwidth}
\end{minipage}
\end{center}
\caption{\it Bar-charts with the total numbers of events in 5/fb at 7~TeV that pass the signal selections of the respective 
searches, showing the relative contributions of the individual classes of subprocesses.}
\label{fig:searchoverview}
\end{figure}

In Figure~\ref{fig:searchoverview} we display the total number of events (Table~\ref{tab:searchoverview}) that pass the selection cuts of the 4 searches in question for the 3 benchmark masses, showing in particular the relative contributions of the 8 classes of production subprocesses.
We see that the numbers of events surviving the $L_p$ single-lepton
cuts are typically $\sim 10$\% of the numbers of events surviving the
$\alpha_T$ MET selection, with much smaller numbers for the
opposite- and same-sign (OS and SS) dilepton searches.
As a consequence, the latter contribute very little to the overall
model sensitivity, which is dominated by the $\alpha_T$ analysis
with significant support only from the $L_p$ analysis.

\subsection{Contributions of Level (2,0) States to MET Searches}

We have already mentioned some reasons why the results in the
realistic model should differ to some extent from those obtained in
the toy model. For one thing, there are additional particles in the
(1,0) tier, beyond the analogue of the $U_1$ quark recurrence
discussed in the toy model, leading to the more complicated decay
chains with leptons discussed above. Secondly, in the realistic model,
higher tiers are present: in particular, the even level (2,0) quarks play an important
role in the full model. They contribute to subprocesses with different
kinematics from the (1,0) tier quarks, giving distributions that differ from those of the toy model.

\begin{figure}
\begin{minipage}{.5\textwidth}
\epsfig{file=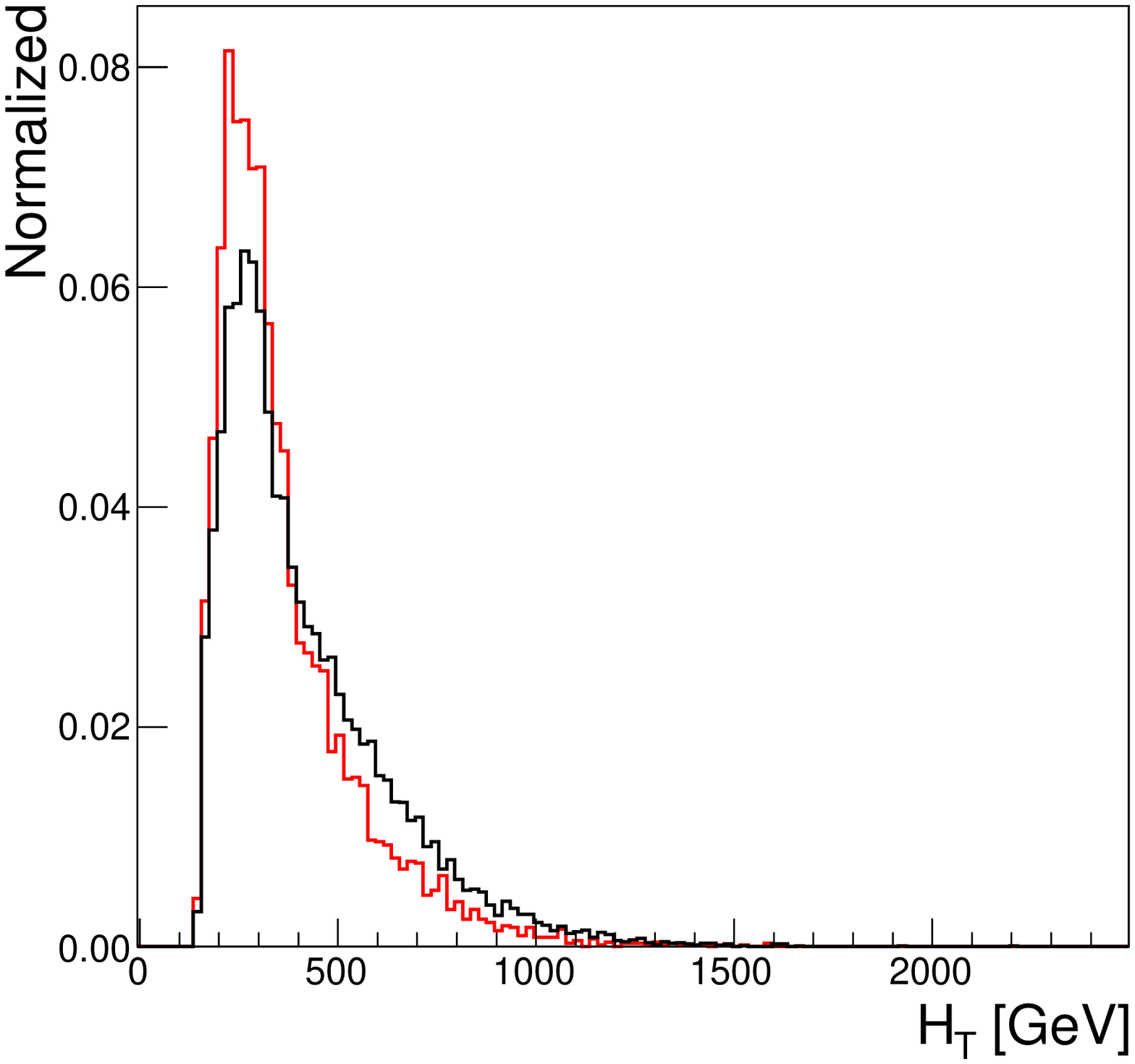,width=\textwidth}
\end{minipage}
\begin{minipage}{.5\textwidth}
\epsfig{file=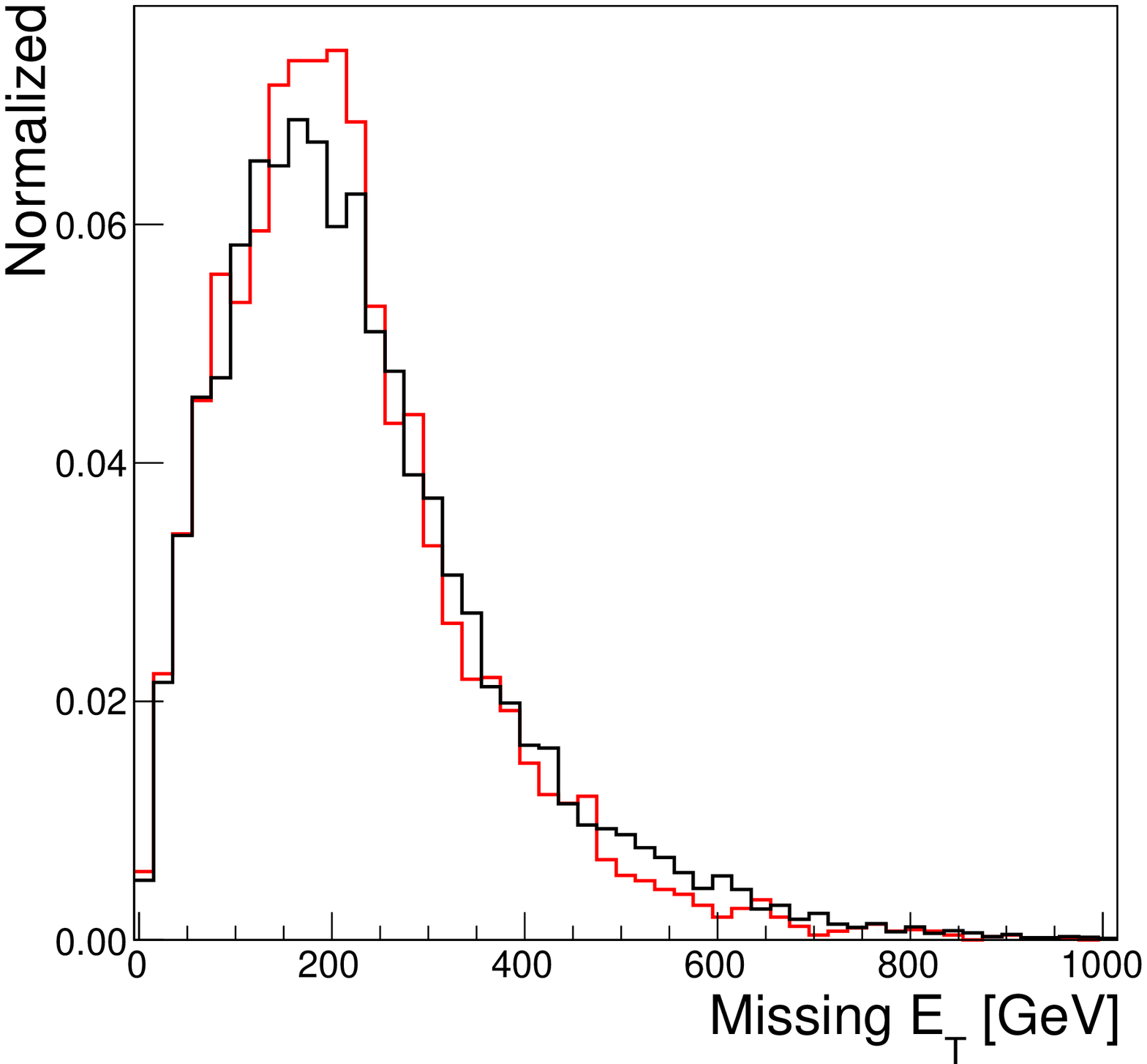,width=\textwidth}
\end{minipage}
\caption{\it The normalised $H_T$ (left) and MET distributions (right) in the toy model  with
$M_{U_1} = 400$~GeV and $\Delta M = 20$~GeV (red) and for the $Q_{1}Q_{1}$ and $Q_{1}\bar{Q_{1}}$ subset of the full model with $m_{KK} = 400$~GeV (black).}
\label{fig:tmvsq10}
\end{figure}

\begin{figure}
\begin{minipage}{.5\textwidth}
\epsfig{file=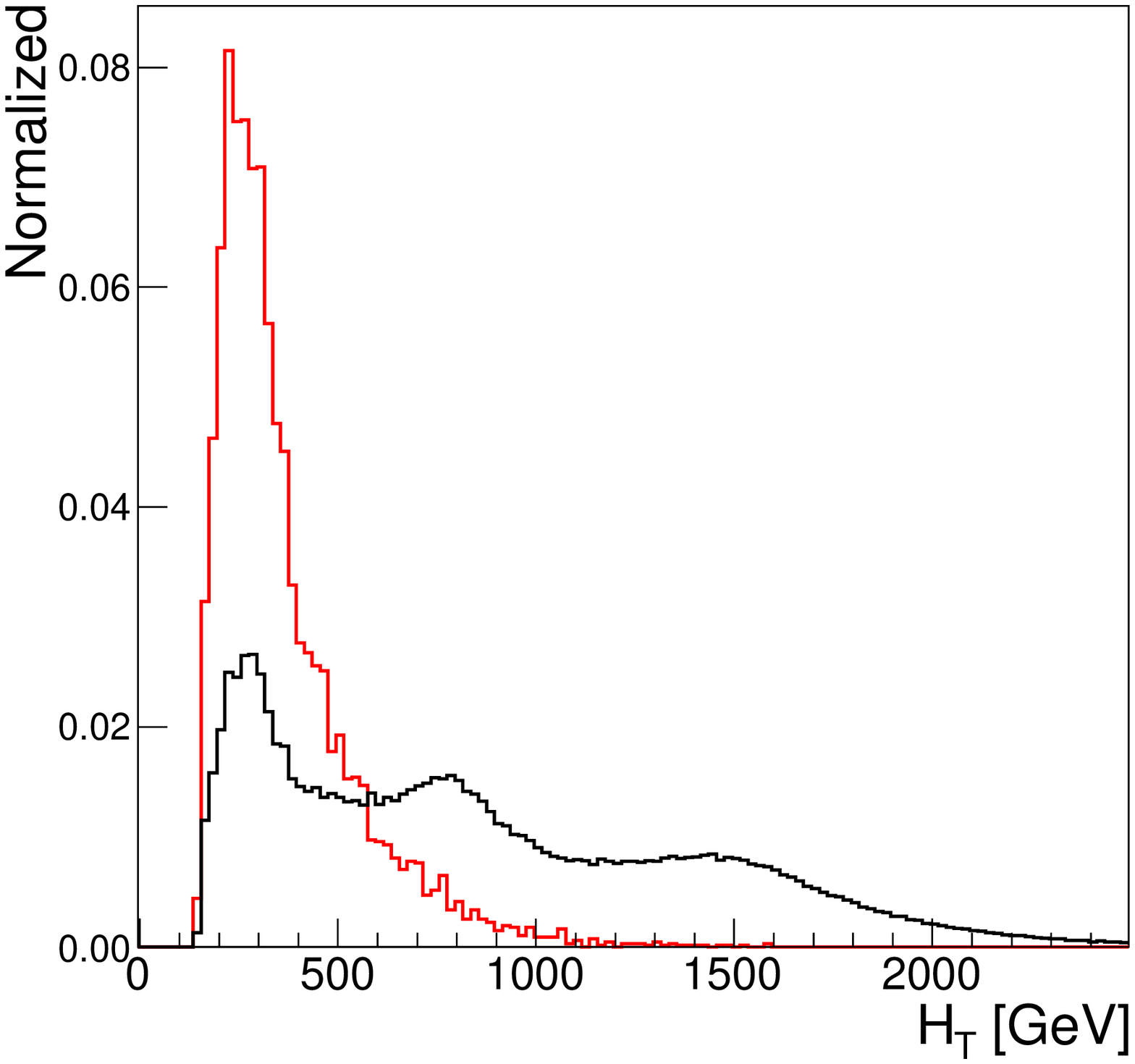,width=\textwidth}
\end{minipage}
\begin{minipage}{.5\textwidth}
\epsfig{file=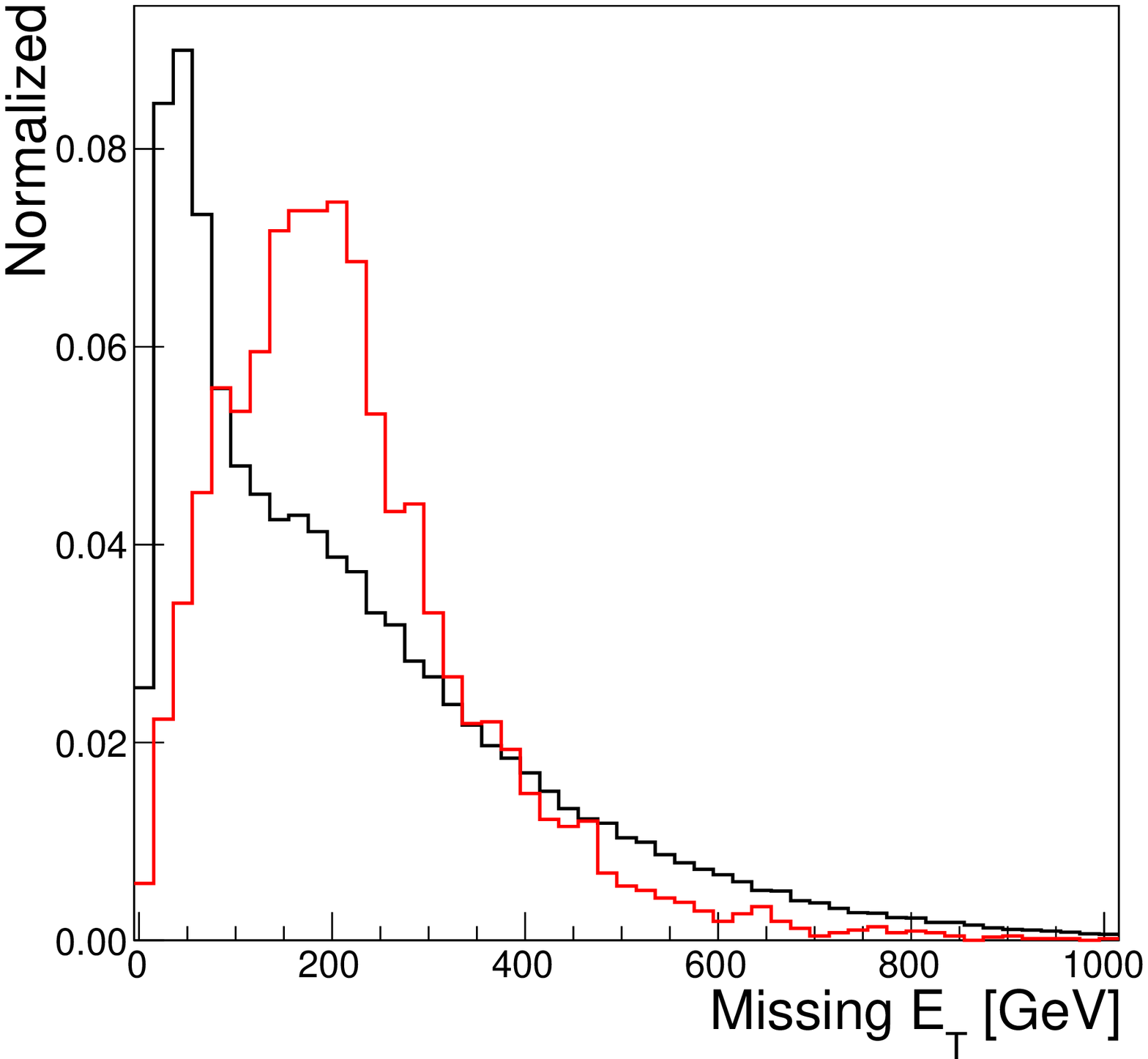,width=\textwidth}
\end{minipage}
\caption{\it The normalised $H_T$ (left) and MET distributions (right) in the toy model with
$M_{U_1} = 400$~GeV and $\Delta M = 20$~GeV (red) and in the full model with $m_{KK} = 400$~GeV (black).
Note the important contributions of the (2,0) states at larger $H_T$ and MET, and the contributions of
more complicated decay chains at smaller MET.}
\label{fig:tmvsfm}
\end{figure}

In order to qualify and quantify the differences, in Figs.~\ref{fig:tmvsq10}
and~\ref{fig:tmvsfm} we display two key kinematical variables used in the
analysis, namely $H_T$ and MET, in the RP$^2$ model for $m_{KK} = 400$~GeV, and in the toy model for $M_{U_1} = 400$~GeV 
and $\Delta M = 20$~GeV. As a first cross-check,
Fig.~\ref{fig:tmvsq10} shows the sum of the $Q_{1}Q_{1}$ and $Q_{1}\bar{Q_{1}}$
subprocesses (black) compared to the toy model histograms (red),
with both distributions normalised to the same number of events. The similarity in the shapes
are as expected, and the comparison validates the toy model approach. When compared to the full
model in Fig.~\ref{fig:tmvsfm}, however, we see a much stronger
extension of the $H_T$ distribution (left) above 300~GeV with a notable bump around 800 GeV, probably coming from resonant hadronic decays of the even (2) gauge bosons. We also see
in the MET distribution (right) a peak at small values but also a
stronger extension above 200~GeV. These extensions of the $H_T$ and
MET distributions are due to the contributions of the second-tier
(2,0) states, as is the low-MET peak. The extensions of the $H_T$
and MET distributions imply that the sensitivities of typical
supersymmetry-motivated MET searches at the LHC to the full model are
greater than those to the toy model, as we now discuss.

\begin{table}[tb]
\begin{tabular}{|c|ccc|ccc|ccc|}
\hline
 & \multicolumn{3}{c|}{Total cross section [fb]} & \multicolumn{3}{c|}{$\alpha_T$ efficiency} & \multicolumn{3}{c|}{$L_p$ efficiency} \\
$m_{KK}$ [GeV]  & 400 & 600 & 700 & 400 & 600 & 700 & 400 & 600 & 700 \\
 \hline
$Q_{1} \bar{Q}_{1}$ & 7440 & 531  & 179  &  0.92\% & 0.91\% & 0.81\% & 0.034\% & 0.032\% & 0.027\%\\
$G_{1} Q_{1}$ & 4780 & 327 & 119 & 0.96\% & 1.27\% & 1.43\% & 0.026\% & 0.039\% & 0.046\%\\
$Q_{1} Q_{1}$ & 1630 & 221 & 99 & 0.80\% & 0.77\% & 0.66\% & 0.027\% & 0.023\%  & 0.022\%\\
$G_{1} G_{1}$ & 418 & 18 & 5.7 & 1.36\% & 1.88\%  & 2.14\% & 0.031\% & 0.05\%  & 0.05\% \\
\hline
$Q_{2} Q_{2}$ & 718 & 25.8 & 5.2 & 6.76\% & 4.76\% & 4.7\% & 2.05\% & 2.63\% & 3.35\% \\
$G_{2} Q_{2}$ & 476 & 7.2 & 1.08 & 5.27\% & 4.36\% & 4.6\% & 1.55\% & 2.0\% & 2.7\% \\
$Q_{2} \bar{Q}_{2}$ & 159 & 2.6 & 0.36 & 5.23\% & 4.32\% & 3.5\% & 1.84\% & 2.84\% & 3.08\% \\
$G_{2} G_{2}$ & 43 & 0.4 & 0.05 & 4.8\% & 4.0\% & 4.1\% & 1.27\% & 1.5\% & 2.0\% \\
 \hline
 total & 15700 & 1130 & 409 & 1.4\% & 1.1\% & 1.0\% & 0.19\% & 0.11\% & 0.08\% \\
 \hline
 \end{tabular}
\caption{\it Cross sections and efficiencies of the $\alpha_T$ and $L_p$ searches in the individual classes, and for
the three recurrence scales considered, $m_{KK} = 400, 600$ and $700$~GeV.} \label{tab:efficiencies}
\end{table}

The impact of the (2,0) states on the searches is also shown in Fig.~\ref{fig:searchoverview}, where we see that, in the case $m_{KK} = 400$~GeV, the even states contribute $\sim 40$\% of the events expected in the $\alpha_T$ MET search and over 80\%
of the events expected in the $L_p$ single-lepton search.
For the $\alpha_T$ search, which contains the largest number of events, the importance of the even states decreases considerably for larger masses, dropping to about 10\% and 5\% for $m_{KK} = 600$ and $700$ GeV.
The main reason behind is the decrease in the production cross sections of a pair of (2) states which is suppressed by a larger mass compared to the (1) states. In fact, the efficiencies of the $\alpha_T$ search in the individual classes of production, listed in Table~\ref{tab:efficiencies}, show a mild decrease in all channels for increasing $m_{KK}$.
On the other hand, in the same Table~\ref{tab:efficiencies} we can see that, while the cross sections for pair production of the odd states at $700$~GeV are between $1\div 5$\% of the values at $400$~GeV,  for the even states the decrease amounts to $0.1\div 1$\%.
The situation is different in the leptonic searches, like $L_p$: in Fig.~\ref{fig:searchoverview} we see that the even tiers are always very important, providing most of the signal events for all values of the masses.
This effect can be understood by looking at the efficiencies of the $L_p$ search in Table~\ref{tab:efficiencies}: while the efficiencies decrease for larger masses in the classes of odd states, for even states the efficiencies increase at larger masses.
This change in efficiencies suffices to compensate for the decrease in cross section.
The increased efficiencies may be explained by the presence of larger MET and leptons with higher $p_T$, 
coming from the decays of the massive $W^\pm_2 \to l^\pm \nu$.

\subsection{Constraints on the UED Scenario}

\begin{table}[tb] \begin{center}
\begin{tabular}{|c|cc|cc|cc|}
\hline
  & \multicolumn{2}{c|}{$m_{KK} = 400$ GeV} & \multicolumn{2}{c|}{$m_{KK} = 600$ GeV} & \multicolumn{2}{c|}{$m_{KK} = 700$ GeV}\\
  & $\epsilon_{\rm total}$ & CL  & $\epsilon_{\rm total}$ & CL  & $\epsilon_{\rm total}$ & CL \\
 \hline
 $\alpha_{T}$ & 1.4\% & 100\% & 1.1\% & 99\% & 1.0\% & 64\% \\
 $L_{p}$ & 0.19\% & 100\% & 0.11\% & 83\% & 0.08\% & 38\%\\
 $OS$ & 0.03\% & 87\% & 0.02\% & 3\% & 0.02\% & 1\%\\
 $SS$ & 0.01\% & 100\% & $<$ 0.01\% & 20\% & $<$ 0.01\% & 5\%\\
\hline
 Combination & & 100\% & & 99.9\% & & 72\% \\
 \hline
 \end{tabular}
\end{center} \caption{\it The efficiencies $\epsilon_{total}$ and the exclusion confidence levels CL
 for the signatures considered in the CMS 7~TeV data: the $\alpha_T$ MET search, the $L_p$ single-lepton
 search, and the opposite- and same-sign (OS and SS) dilepton searches, as well as the combined CL.
 Results are given for the three recurrence scales considered, $m_{KK} = 400, 600$ and $700$~GeV.}
 \label{tab:results}
 \end{table}

The results of the previous sections can be used to test the reach of the 7~TeV searches on the RP$^2$ UED model.
Here we only use the results of CMS, in particular the $\alpha_T$~\cite{Chatrchyan:2011zy}, the $L_p$ single lepton~\cite{CMSLpsearch}, and same sign~\cite{CMSSSsearch} and opposite sign~\cite{CMSOSsearch} (SS and OS) searches.
The number of events that pass the selection cuts, listed in Table~\ref{tab:searchoverview}, can be combined with the experimental results to obtain a bound.
We use the CL$_s$ statistic to calculate
the exclusion confidence levels (CL) shown in Table~\ref{tab:results} for $m_{KK} = 400, 600$
and 700~GeV, respectively. Also shown is the exclusion confidence level obtained by
combining the different searches for each value of $m_{KK}$.

We see that the case $m_{KK} = 400$~GeV is excluded with a very high degree of confidence independently
by the $\alpha_T$, $L_p$ and SS analyses, and disfavoured to a lesser extent by the OS analysis. This value
of $m_{KK}$ is therefore excluded very robustly. The case $m_{KK} = 600$~GeV is excluded at the
99\% CL by the $\alpha_T$ analysis alone, and at the 83\% by the $L_p$ analysis, but not significantly by
the dilepton analyses, which start losing reach at larger masses. Combining the analyses, we find that this value of $m_{KK}$ is also excluded
robustly, at the 99.9\% CL. However, the case $m_{KK} = 700$~GeV cannot be excluded: it is disfavoured
at the 64\% CL by the $\alpha_T$ analysis and at the 38\% CL by the $L_p$ analysis, but the combined
CL is only 72\%, with negligible contributions from the OS and SS analyses.

We can therefore conclude that the LHC MET searches (CMS) at 7~TeV require a bound on the $m_{KK}$ mass in the RP$^2$ model between $600$ and $700$~GeV.
It is interesting to compare this number with the expected bound from searches without MET: the most promising one is the bound on resonant di-leptons originating from the decays of $A_2$ and $Z_2$, which are produced in the decay chain of even coloured states.
Even though the events contain other particles, and the leptons are coming from two resonances, the number of events can be directly compared to the exclusion for a single $Z' \to l^+ l^-$. It was shown in~\cite{Cacciapaglia:2012dy}
that the corresponding bound is $m_{KK} > 575$~GeV, and other channels such as $W^\pm_2 \to l^\pm \nu$
and jet resonances are less competitive.
This comparison shows that searches with MET are already competitive
with searches that do not require the presence of MET, and seem already to have a better reach.
Nevertheless, a possibly stronger bound might be obtained considering searches in pair di-jet final states: in the RP$^2$ such signal can be originated
by pair-produced (2,0) states that decay directly into SM. The analysis of this signal will be undertaken in a forthcoming study.

\section{Improving the LHC Sensitivity to the UED Scenario}

We display in Tab.~\ref{tab:efficiencies} the essential problem that
must be confronted to improve the sensitivity of an LHC experiment such as CMS to the UED
scenario. For all the values of $m_{KK}$ studied, the
largest cross section for producing Kaluza-Klein excitations is that for ${\bar Q_1} Q_1$
production, whereas the cross sections for producing a second-tier quark excitation together with a gluon
excitation and/or another second-tier quark excitation are much smaller.
Looking at the distributions in Figs.~\ref{fig:tmvsq10} and \ref{fig:tmvsfm},
we see that the pair production of odd states have similar distributions as the toy model, 
reflecting the similar low efficiencies for compressed spectra displayed in Tab.~\ref{tab:efficiencies}.

It is clear that the LHC experiments should strive to increase their efficiencies for the
channels with the largest cross sections, particularly ${\bar Q_1} Q_1$ production.
As already noted, we have found the sensitivities of the ATLAS $m_{eff}$ search and
monojet searches to be less than that of the CMS $\alpha_T$ search. Therefore, the latter
may provide the best starting-point for the optimisation of the UED search. We recall
that this search was originally motivated by the search for supersymmetry, though the
kinematical philosophy is applicable to any scenario with a strong hierarchy of masses
and a massive, invisible dark
matter particle. There may be scope for
improving the efficiency for the UED scenario by varying the jet energy and rapidity
analyses used in the $\alpha_T$ analysis, e.g., taking into account the fact that many
of the jets in the UED scenario are due to initial-state radiation. Improving the efficiencies for such channels
would also benefit searches for supersymmetric models with compressed spectra.
However, a study of this possibility requires understanding the experimental capabilities in more detail,
and lies beyond the scope of this work.

On the other hand, we also see in Figs.~\ref{fig:tmvsq10} and \ref{fig:tmvsfm} that the
even tiers have signatures with very different distributions
from the toy model, that are peculiar to UED models.
As we see in Tab.~\ref{tab:efficiencies} the efficiencies for the most 
relevant $\alpha_T$ and $L_p$ searches for $m_{KK} = 400, 600$ and 700~GeV
are relatively large for final states with (2,0) states,
and are quite similar in all these channels for all the recurrence scales studied. These
relatively high efficiencies are to be compared with the much lower efficiencies for
the ${\bar Q_1} Q_1$ and other channels that have large cross sections.

\begin{figure}
\begin{minipage}{.5\textwidth}
\epsfig{file=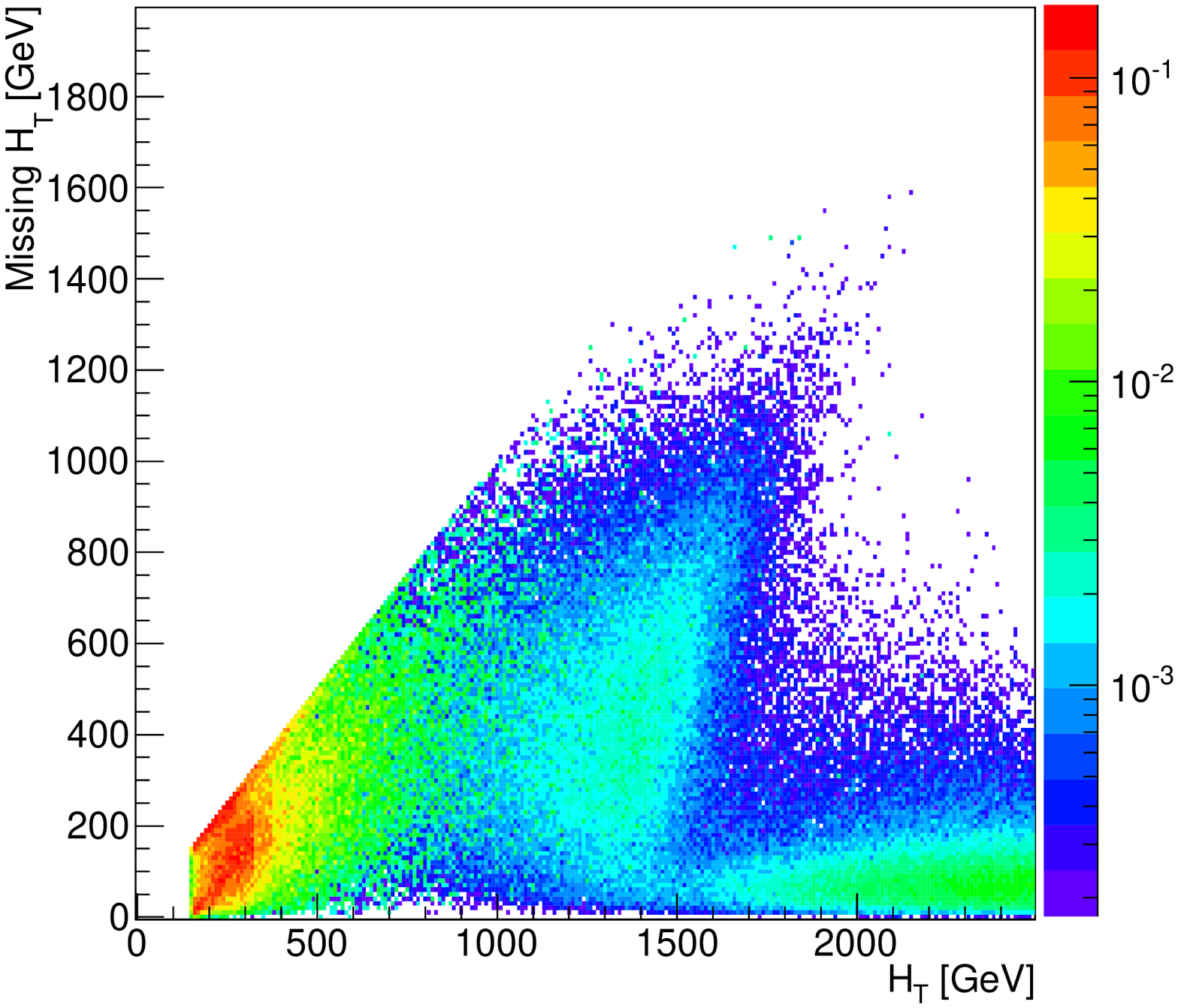,width=\textwidth}
\end{minipage}
\begin{minipage}{.5\textwidth}
\epsfig{file=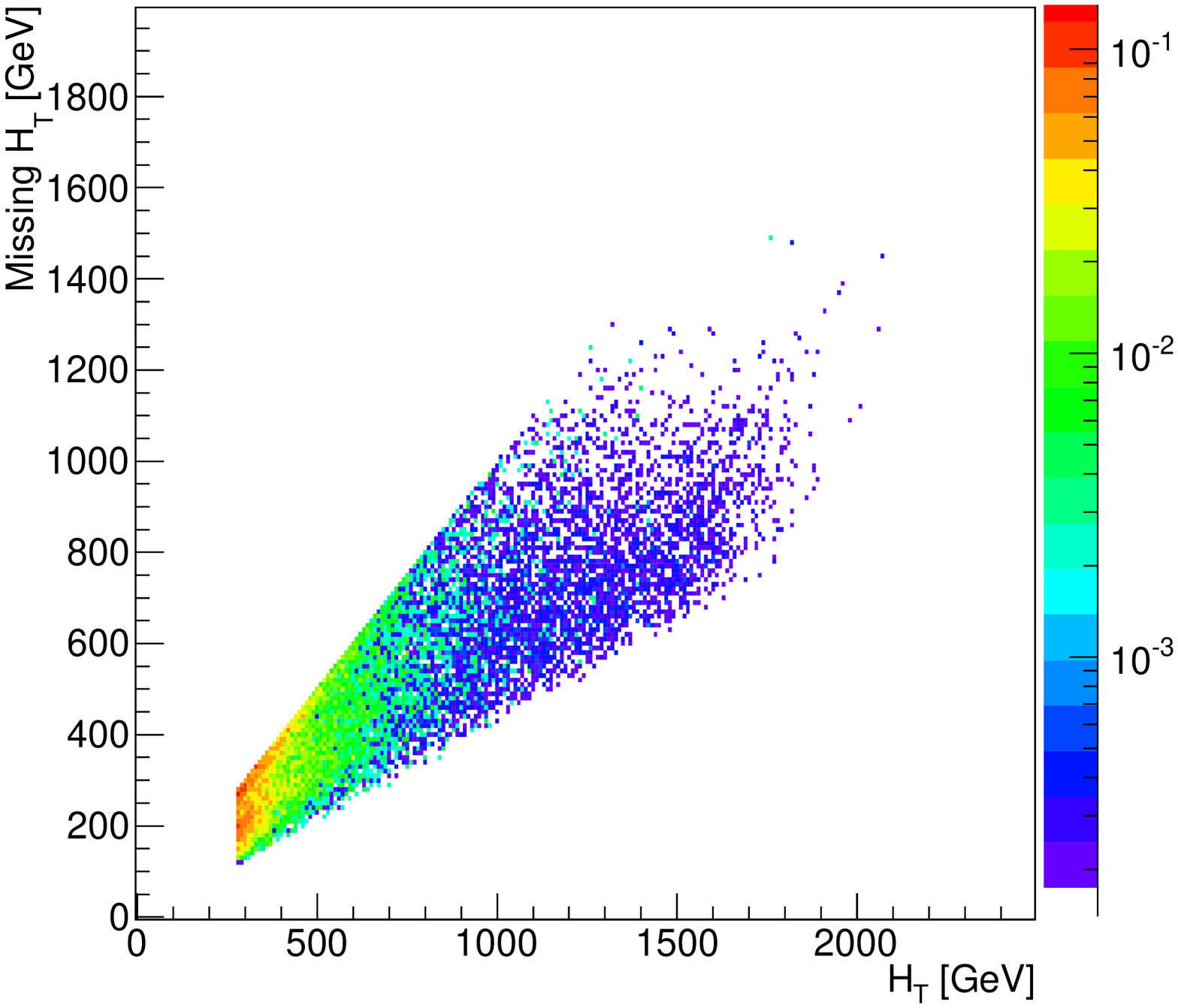,width=\textwidth}
\end{minipage}
\caption{
{\it
Scatter plots for the UED model with $m_{KK} = 700$~GeV in the $(H_T, MH_T)$ plane
(left) before applying the CMS $\alpha_T$ cut and (right) afterwards. The distribution is normalised to 1.}}
\label{fig:HTvsMHT} 
\end{figure}

We recall that the cross sections for producing tier-2 states will be larger at
8~TeV and higher LHC energies than they were at 7~TeV, 
and that the increase will be more important for the even tier than for the lighter odd tier.
Therefore, in designing an optimal strategy for the UED model it is natural
to focus attention on the tier-2 states, with the aim of improving the reach
by adapting the cuts that were optimised for searches for supersymmetric
models that do not contain analogous states. As an aid to this effort,
in Fig.~\ref{fig:HTvsMHT} we display scatter plots in the $(H_T, MH_T)$ plane
for the UED model with $m_{KK} = 700$~GeV 
before applying the CMS $\alpha_T$ cut (in the left panel) and after applying it (in the right panel).
We see that the $\alpha_T$ cut, which was largely motivated by supersymmetric models
with hierarchical masses and corresponding MET signatures, removes large clusters of UED events with relatively
large values of $H_T/MH_T$, notably around $H_T \sim 1400$~GeV and $MH_T \sim 400$~GeV
and $H_T \sim 2200$~GeV and $MH_T \sim 100$~GeV.
However, a study of the cut optimisation again requires understanding the experimental capabilities in more detail,
and lies beyond the scope of this work.

\section{Summary and Conclusions}

In this paper we have explored the sensitivity to scenarios with extra dimensions of the LHC MET searches,
which were motivated initially by searches for supersymmetry. Like the more-studied supersymmetric
scenarios, the extra-dimensional scenarios we study have a Dark matter candidate, here in the form of
the lightest Kaluza-Klein recurrence. However, the MET signal is in general reduced compared to
generic supersymmetric scenarios, because the splitting between the masses of different states in the same tier are relatively
small. We have used {\tt MadGraph} to calculate the signals to be expected in a toy model
with a single Kaluza-Klein quark recurrence, which we interfaced
with {\tt PYTHIA} and simulated for the ATLAS and CMS detectors using an implementation of {\tt Delphes} that
had been validated previously for supersymmetry searches. 

We then presented a more realistic model with two extra dimensions, and showed how the full set of
first-level Kaluza-Klein recurrences would yield events with final-state leptons as well as MET events.
We also showed that second-level Kaluza-Klein recurrences would make important additional
contributions to the MET signals. As a result of these two additional signatures, the sensitivity of
current LHC MET searches to the realistic model is greater than to the toy model,
reaching into the range of masses where the lightest Kaluza-Klein recurrence could realistically
provide all the cosmological dark matter. Specifically, in this first study we find that a Kaluza-Klein recurrence
scale of 600~GeV is robustly excluded at the 99.9\% CL, whereas a scale of 700~GeV is
disfavoured only at the 72\% CL. These sensitivities are due principally to a combination of the
CMS $\alpha_T$ and single-lepton $L_p$ analyses, with opposite- and same-sign dilepton
searches playing a much less important r\^ole. It will be interesting to extend the current study to the
full LHC 8-TeV data set as soon as sufficient details of the MET searches are made available for
calculating their sensitivity to extra-dimensional scenarios to be feasible.
This study also showed that the reach of searches with MET already
extends beyond the reach of dilepton resonance searches at 7~TeV, 
which give a bound of $575$~GeV~\cite{Cacciapaglia:2012dy}.

Studies of the LHC sensitivity to extra-dimensional scenarios are still much less developed than
the corresponding supersymmetry studies, and much work remains to be done. For example,
one potentially important class of signatures that we did not explore is
$pp \to X_2$ production followed by $X_2 \to X_1 X_1, \, X_2 x_{SM}$ decays, where $X_{1,2}$ denote generic first- and
second-level Kaluza-Klein recurrences. The resonant production of the $X_2$ state takes place 
via a loop-induced vertex, and the calculation and simulation of this class of signal would be an
interesting next step beyond the present work. We also note that there are several refinements
of the signatures studied in this work that could be implemented, such as a more complete
calculation of the spectra for different ratios of the sizes of the two extra dimensions, as well as
an accurate treatment of spin effects and correlations.

However, we think that the work presented here provides the LHC experiments with a
benchmark that they could use in making their own interpretations of MET searches in the
context of models with extra dimensions. In many ways, the minimal RP$^2$ scenario
studied here is simpler than the CMSSM, for example, with fewer parameters. On the other
hand, our study has demonstrated the need for a complete analysis including higher-level
Kaluza-Klein recurrences that have no analogue in supersymmetric models. One possible
use of this benchmark model would be consider the design of LHC MET searches to be
performed at higher energies after the shutdown, so that they are optimised for a broader
class of models than popular supersymmetric scenarios.

To paraphrase Shakespeare, {\it ``There may more things at the LHC, ATLAS and CMS,
than are dreamt of in your current trigger, event selection and analysis philosophy".}

\section*{Acknowledgements}

We would like to thank O. Buchmueller for helpful discussions. 
The work of JE and JM was supported partly by the London
Centre for Terauniverse Studies (LCTS), using funding from the European
Research Council via the Advanced Investigator Grant 267352.


\begin{thebibliography}{99}


\bibitem{LHCsearches}
A complete list of searches can be found on the Supersymmetry Physics Group Twiki page of ATLAS: \\
\href{https://twiki.cern.ch/twiki/bin/view/AtlasPublic/SupersymmetryPublicResults}{\tt https://twiki.cern.ch/twiki/bin/view/AtlasPublic/SupersymmetryPublicResults} and of CMS: \\
\href{https://twiki.cern.ch/twiki/bin/view/AtlasPublic/SupersymmetryPublicResults}{\tt https://twiki.cern.ch/twiki/bin/view/CMSPublic/PhysicsResultsSUS}.

\bibitem{LeCompte:2011fh}
  T.~J.~LeCompte and S.~P.~Martin,
  Phys.\ Rev.\ D {\bf 85} (2012) 035023
  [\href{http://arxiv.org/abs/1111.6897}{arXiv:1111.6897} [hep-ph]].

\bibitem{Dreiner:2012gx}
  H.~K.~Dreiner, M.~Kramer and J.~Tattersall,
  Europhys.\ Lett.\  {\bf 99} (2012) 61001
  [\href{http://arxiv.org/abs/1207.1613}{arXiv:1207.1613} [hep-ph]].

\bibitem{Bhattacherjee:2012mz}
  B.~Bhattacherjee and K.~Ghosh,
  [\href{http://arxiv.org/abs/1207.6289}{arXiv:1207.6289} [hep-ph]].
  
\bibitem{Antoniadis:1990ew}
  I.~Antoniadis,
  Phys.\ Lett.\ B {\bf 246} (1990) 377.

\bibitem{Appelquist:2000nn}
  T.~Appelquist, H.~C.~Cheng and B.~A.~Dobrescu,
  Phys.\ Rev.\  D {\bf 64} (2001) 035002
  [\href{http://arxiv.org/abs/hep-ph/0012100}{arXiv:hep-ph/0012100}].
  
\bibitem{LKP}
  G.~Servant and T.~M.~P.~Tait,
  Nucl.\ Phys.\  B {\bf 650}, 391 (2003)
  [\href{http://arxiv.org/abs/hep-ph/0206071}{arXiv:hep-ph/0206071}].

\bibitem{ArkaniHamed:2002qx}
  N.~Arkani-Hamed, A.~G.~Cohen, E.~Katz, A.~E.~Nelson, T.~Gregoire and J.~G.~Wacker,
  JHEP {\bf 0208} (2002) 021
  [\href{http://arxiv.org/abs/hep-ph/0206020}{hep-ph/0206020}].

\bibitem{ArkaniHamed:2002qy}
  N.~Arkani-Hamed, A.~G.~Cohen, E.~Katz and A.~E.~Nelson,
  JHEP {\bf 0207} (2002) 034
  [\href{http://arxiv.org/abs/hep-ph/0206021}{hep-ph/0206021}].

\bibitem{ArkaniHamed:2001ca}
  N.~Arkani-Hamed, A.~G.~Cohen and H.~Georgi,
  Phys.\ Rev.\ Lett.\  {\bf 86} (2001) 4757
  [\href{http://arxiv.org/abs/hep-ph/0104005}{hep-th/0104005}].

\bibitem{Low:2004xc}
  I.~Low,
  JHEP {\bf 0410} (2004) 067
  [\href{http://arxiv.org/abs/hep-ph/0409025}{arXiv:hep-ph/0409025}].

\bibitem{Bhattacherjee:2010vm}
  B.~Bhattacherjee and K.~Ghosh,
  Phys.\ Rev.\ D {\bf 83} (2011) 034003
  [\href{http://arxiv.org/abs/1006.3043}{arXiv:1006.3043} [hep-ph]].

\bibitem{Belyaev:2012ai}
  A.~Belyaev, M.~Brown, J.~Moreno and C.~Papineau,
  [\href{http://arxiv.org/abs/1212.4858}{arXiv:1212.4858} [hep-ph]].
  
\bibitem{Perelstein:2011ds}
  M.~Perelstein and J.~Shao,
  Phys.\ Lett.\ B {\bf 704} (2011) 510
  [\href{http://arxiv.org/abs/1103.3014}{arXiv:1103.3014} [hep-ph]].

\bibitem{Cacciapaglia:2009pa}
  G.~Cacciapaglia, A.~Deandrea and J.~Llodra-Perez,
  JHEP {\bf 1003} (2010) 083
  [\href{http://arxiv.org/abs/0907.4993}{arXiv:0907.4993} [hep-ph]].

\bibitem{Arbey:2012ke}
  A.~Arbey, G.~Cacciapaglia, A.~Deandrea and B.~Kubik,
  JHEP {\bf 1301} (2013) 147
  [\href{http://arxiv.org/abs/1210.0384}{arXiv:1210.0384} [hep-ph]].

\bibitem{Alwall:2011uj}
  J.~Alwall, M.~Herquet, F.~Maltoni, O.~Mattelaer and T.~Stelzer,
  JHEP {\bf 1106} (2011) 128
  [\href{http://arxiv.org/abs/1106.0522}{arXiv:1106.0522} [hep-ph]].
  
\bibitem{Cacciapaglia:2012dy}
  G.~Cacciapaglia and B.~Kubik,
  JHEP {\bf 1302} (2013) 052
  [\href{http://arxiv.org/abs/1209.6556}{arXiv:1209.6556} [hep-ph]].

\bibitem{Cacciari:2011hy}
  M.~Cacciari, M.~Czakon, M.~Mangano, A.~Mitov and P.~Nason,
  Phys.\ Lett.\ B {\bf 710} (2012) 612
  [\href{http://arxiv.org/abs/1111.5869}{arXiv:1111.5869} [hep-ph]].

\bibitem{Cacciari_ttbar}
M. Cacciari, M. Czakon, M.L. Mangano, A. Mitov and P. Nason, \\
  \href{http://www.lpthe.jussieu.fr/~cacciari/ttbar/}{\tt http://www.lpthe.jussieu.fr/~cacciari/ttbar/}

\bibitem{Sjostrand:2006za}
  T.~Sjostrand, S.~Mrenna and P.~Z.~Skands,
  JHEP {\bf 0605} (2006) 026
  [\href{http://arxiv.org/abs/hep-ph/0603175}{hep-ph/0603175}].

\bibitem{Ovyn:2009tx}
  S.~Ovyn, X.~Rouby and V.~Lemaitre,
  [\href{http://arxiv.org/abs/0903.2225}{arXiv:0903.2225} [hep-ph]].

\bibitem{Alwall:2008qv}
  J.~Alwall, S.~de Visscher and F.~Maltoni,
  JHEP {\bf 0902} (2009) 017
  [\href{http://arxiv.org/abs/0810.5350}{arXiv:0810.5350} [hep-ph]].
  
\bibitem{MC8}
  O.~Buchmueller {\it et al.},
  Eur.\ Phys.\ J.\ C {\bf 72} (2012) 2243
  [\href{http://arxiv.org/abs/1207.7315}{arXiv:1207.7315} [hep-ph]].
  
\bibitem{Allanach:2001kg}
  B.~C.~Allanach,
  Comput.\ Phys.\ Commun.\  {\bf 143} (2002) 305
  [\href{http://arxiv.org/abs/hep-ph/0104145}{hep-ph/0104145}].
    
\bibitem{ATLAS5fb}
The ATLAS collaboration,
47th Rencontres de Moriond on QCD and High Energy Interactions, La Thuile, Italy, 10 - 17 Mar 2012, 
\href{https://cdsweb.cern.ch/record/1432199}{\tt https://cdsweb.cern.ch/record/1432199}.

\bibitem{CMSmT2}
CMS collaboration,\\
\href{https://twiki.cern.ch/twiki/bin/view/CMSPublic/PhysicsResultsSUS12002}{\tt https://twiki.cern.ch/twiki/bin/view/CMSPublic/PhysicsResultsSUS12002}.

\bibitem{CMSrazor}
CMS collaboration,\\
\href{https://twiki.cern.ch/twiki/bin/view/CMSPublic/PhysicsResultsSUS12005}{\tt https://twiki.cern.ch/twiki/bin/view/CMSPublic/PhysicsResultsSUS12005}.

\bibitem{Chatrchyan:2011zy}
  S.~Chatrchyan {\it et al.}  [CMS Collaboration],
  Phys.\ Rev.\ Lett.\  {\bf 107} (2011) 221804
  [\href{http://arxiv.org/abs/1109.2352}{arXiv:1109.2352} [hep-ex]].
  
\bibitem{Alwall:2008ag}
  J.~Alwall, P.~Schuster and N.~Toro,
  Phys.\ Rev.\ D {\bf 79} (2009) 075020
  [\href{http://arxiv.org/abs/0810.3921}{arXiv:0810.3921} [hep-ph]].
  
\bibitem{SMS}
S.~Chatrchyan {\it et al.}  [CMS Collaboration],
  [\href{http://arxiv.org/abs/1301.2175}{arXiv:1301.2175} [hep-ex]].
  
\bibitem{Chatrchyan:2012me}
  S.~Chatrchyan {\it et al.}  [CMS Collaboration],
  JHEP {\bf 1209} (2012) 094
  [\href{http://arxiv.org/abs/1206.5663}{arXiv:1206.5663} [hep-ex]].
  
\bibitem{Dobrescu:2004zi}
  B.~A.~Dobrescu and E.~Ponton,
  JHEP {\bf 0403} (2004) 071
  [\href{http://arxiv.org/abs/hep-ph/0401032}{arXiv:hep-th/0401032}].

\bibitem{Cheng:2002iz}
  H.~C.~Cheng, K.~T.~Matchev and M.~Schmaltz,
  Phys.\ Rev.\  D {\bf 66}, 036005 (2002)
  [\href{http://arxiv.org/abs/hep-ph/0204342}{arXiv:hep-ph/0204342}].

\bibitem{Ponton:2005kx}
  E.~Ponton and L.~Wang,
  JHEP {\bf 0611} (2006) 018
  [\href{http://arxiv.org/abs/hep-ph/0512304}{arXiv:hep-ph/0512304}].

\bibitem{Belanger:2010yx}
  G.~Belanger, M.~Kakizaki and A.~Pukhov,
  JCAP {\bf 1102} (2011) 009
  [\href{http://arxiv.org/abs/1012.2577}{arXiv:1012.2577} [hep-ph]].

\bibitem{Dobrescu:2007ec}
  B.~A.~Dobrescu, D.~Hooper, K.~Kong and R.~Mahbubani,
  JCAP {\bf 0710} (2007) 012
  [\href{http://arxiv.org/abs/0706.3409}{arXiv:0706.3409} [hep-ph]].

\bibitem{Dohi:2010vc}
  H.~Dohi and K.~-y.~Oda,
  Phys.\ Lett.\ B {\bf 692} (2010) 114
  [\href{http://arxiv.org/abs/1004.3722}{arXiv:1004.3722} [hep-ph]].

\bibitem{Cacciapaglia:2011hx}
  G.~Cacciapaglia, A.~Deandrea and J.~Llodra-Perez,
  JHEP {\bf 1110} (2011) 146
  [\href{http://arxiv.org/abs/1104.3800}{arXiv:1104.3800} [hep-ph]].

\bibitem{Cacciapaglia:2011kz}
  G.~Cacciapaglia, R.~Chierici, A.~Deandrea, L.~Panizzi, S.~Perries and S.~Tosi,
  JHEP {\bf 1110} (2011) 042
  [\href{http://arxiv.org/abs/1107.4616}{arXiv:1107.4616} [hep-ph]].
  
\bibitem{Christensen:2008py}
  N.~D.~Christensen and C.~Duhr,
  [\href{http://arxiv.org/abs/0806.4194}{arXiv:0806.4194} [hep-ph]].

\bibitem{Pukhov:1999gg}
  A.~Pukhov, E.~Boos, M.~Dubinin, V.~Edneral, V.~Ilyin, D.~Kovalenko, A.~Kryukov and V.~Savrin {\it et al.},
  [\href{http://arxiv.org/abs/hep-ph/9908288}{hep-ph/9908288}];

  A.~Pukhov,
  [\href{http://arxiv.org/abs/hep-ph/0412191}{hep-ph/0412191}].

\bibitem{Meade:2007js}
  P.~Meade and M.~Reece,
  [\href{http://arxiv.org/abs/hep-ph/0703031}{hep-ph/0703031}].

\bibitem{CMSLpsearch}
  S.~Chatrchyan {\it et al.}  [CMS Collaboration],
  [\href{http://arxiv.org/abs/1212.6428}{arXiv:1212.6428} [hep-ex]].

\bibitem{CMSOSsearch}
  S.~Chatrchyan {\it et al.}  [CMS Collaboration],
  Phys.\ Lett.\ B {\bf 718} (2013) 815
  [\href{http://arxiv.org/abs/1206.3949}{arXiv:1206.3949} [hep-ex]].

\bibitem{CMSSSsearch}
  S.~Chatrchyan {\it et al.}  [CMS Collaboration],
  Phys.\ Rev.\ Lett.\  {\bf 109} (2012) 071803
  [\href{http://arxiv.org/abs/1205.6615}{arXiv:1205.6615} [hep-ex]].

\bibitem{Rolbiecki:2012gn}
  K.~Rolbiecki and K.~Sakurai,
  JHEP {\bf 1210} (2012) 071
  [\href{http://arxiv.org/abs/1206.6767}{arXiv:1206.6767} [hep-ph]].


\end{thebibliography}
\end{document}